\documentclass[aps,prd,twocolumn,showpacs,10pt,superscriptaddress,preprintnumbers,nofootinbib]{revtex4-2}
\pdfoutput=1
\usepackage{mathtools}
\newcommand{\fsl}[1]{\ensuremath{\mathrlap{\!\not{\phantom{#1}}}#1}}
\usepackage{subcaption}
\usepackage{amsmath}
\usepackage{amssymb}
\usepackage{multirow}
\allowdisplaybreaks
\usepackage{hyperref}
\allowdisplaybreaks
\hypersetup{
colorlinks=true,
linkcolor=blue,
linktocpage=true,
citecolor=violet
}

\usepackage{xargs}
\usepackage[colorinlistoftodos,prependcaption,textsize=tiny]{todonotes}

\newcommand{\liverpool}{Department of Mathematical Sciences, University of Liverpool, Liverpool L69 3BX, 
U.K.
}

\begin{document}

\title{
Helicity amplitudes in massless QED to higher orders in the dimensional regulator}

\author{Thomas Dave}
\email{tcdave@liverpool.ac.uk}
\affiliation{\liverpool}
\author{William J. Torres~Bobadilla}
\email{torres@liverpool.ac.uk}
\affiliation{\liverpool}

\begin{abstract}
We analytically calculate one- and two-loop helicity amplitudes in massless QED, by adopting a four-dimensional tensor decomposition. 
We draw our attention to 
four-fermion and Compton scattering processes 
to higher orders in the dimensional regulator, as required for theoretical predictions at N$^3$LO.
We organise loop amplitudes by proposing an efficient algorithm at integrand level 
to group Feynman graphs into integral families. 
We study the singular structure of these amplitudes and discuss the correspondence between QED and QCD processes. 
We present our results in terms of generalised polylogarithms up to transcendental weight six.

\end{abstract}

\maketitle

\section{Introduction}

Precision physics plays an increasingly important role in research at particle colliders. 
In recent years, there have been many developments in multi-loop calculation techniques needed to compute scattering amplitudes at increasing perturbative orders in Quantum Field Theories, where their expansion parameter exactly corresponds to the strong coupling constant $\alpha_\text{S}$ in 
Quantum Chromodynamics (QCD) and the fine structure constant $\alpha$ in Quantum Electrodynamics (QED)~\cite{Heinrich:2020ybq,Huston:2023ofk}.

Many efforts have been put in efficiently calculating scattering amplitudes of QCD processes to the highest possible perturbative order~\cite{Andersen:2024czj}. 
In the current frontier, theoretical predictions for collider experiments at high energies are aiming at reaching the best accuracy. 
After the successful automation of tree-level and one-loop scattering amplitude computations, 
which led to the next-to-leading order (NLO) revolution, 
the attention has now shifted towards developing innovative strategies for calculating 
multi-loop scattering amplitudes to enable theoretical predictions beyond NLO (NNLO and beyond).

Focusing on QCD theoretical predictions for four-parton scattering at NNLO~\cite{Anastasiou:2000kg,Anastasiou:2000ue,Anastasiou:2001sv,Anastasiou:2002zn} 
and N$^3$LO~\cite{Ahmed:2019qtg,Caola:2021rqz,Caola:2022dfa,Caola:2021izf}, 
we observe a clear approach, especially at N$^3$LO, for the analytic evaluation of scattering amplitudes required  at this perturbative order. 
By employing the method of differential equations~\cite{Kotikov:2021tai,Gehrmann:1999as,Henn:2013pwa} alongside integral-level relations satisfied by Feynman integrals~\cite{Chetyrkin:1981qh,Laporta:2000dsw},
these results have become accessible.

Interestingly, less attention has been devoted to the calculation of similar processes that appear in QED. 
The construction of those massless scattering amplitudes up to two loops was indeed performed a long time ago~\cite{Bern:2000ie,Bern:2001dg,Anastasiou:2002zn} but to date, 
there has not been an extension of these results. 
Moreover, recent attention has been given to constructing these two-loop scattering amplitudes by keeping the 
dependence on the lepton masses~\cite{CarloniCalame:2020yoz,
Banerjee:2021mty,Bonciani:2021okt,Budassi:2021twh,Broggio:2022htr,Kollatzsch:2022bqa,Badger:2023xtl,Delto:2023kqv,Fadin:2023phc}, motivated by studying physics at low energies~\cite{Banerjee:2020tdt,Aliberti:2024fpq}. 
The presence of the lepton masses certainly increases the complexity in the evaluation of the constituents 
of the amplitudes, the Feynman integrals, whose efficient numerical evaluation is still an undergoing problem~\cite{Bourjaily:2022bwx}.

In this work, we are interested in extending the results of Refs.~\cite{Bern:2000ie,Bern:2001dg,Anastasiou:2002zn} by analytically calculating 
massless QED scattering amplitudes needed for N$^3$LO theoretical predictions. Owing to the dimensional regularisation, we begin our work by calculating, from a diagrammatic approach, 
one- and two-loop scattering amplitudes to higher orders in the dimensional parameter $\epsilon=(4-D)/2$. 
Unlike previous calculations in QED, we adopt a tensor decomposition of scattering amplitudes,
this immediately allows us to express helicity amplitudes and loop interference terms. 

Because in a tensor decomposition of a scattering amplitude 
the number of tensor structures grows according to the loop order when considering arbitrary space-time dimensions, 
we profit from the information that external particles live in four space-time dimensions. 
Specifically, for a given physical process, this allows us to have the number of tensor structures equal to the 
number of independent helicity states, regardless of the loop order~\cite{Chen:2019wyb,Peraro:2019cjj,Peraro:2020sfm}. 
Thus, for the construction of loop amplitudes, 
we opt to use the 't Hooft-Veltman dimensional regularisation scheme~\cite{tHooft:1972tcz} 
to regulate ultraviolet (UV) and infrared (IR) divergences. 
This requires us to consider internal loop momenta in $D$ dimensions~\cite{Gnendiger:2017pys}. 
We remark that any discrepancy between regularisation schemes is overcome 
once IR and UV divergences are removed in the construction of a finite remainder, 
since additional terms that behave as $D-4$ vanish in the four-dimensional limit. 

Due to the number of Feynman diagrams that start appearing at two loops (and the subsequent computation at three loops~\cite{Dave:2024xxx}),
we establish a path to organise our calculation, aimed at minimising the required computations. 
In particular, we provide a method for grouping Feynman diagrams into integral families through matrix transformations,
with the motivation of performing operations at integrand level instead of integral level. 
Once we have our independent integrand families, we proceed by calculating integration-by-parts identities 
to express our form factors in terms of master integrals that admit a $d\log$ representation~\cite{Henn:2020lye,Henn:2021aco}. 
The organisation of our form factors in terms of these master integrals facilitates the computation of the 
complete scattering amplitude, as possible crossings of kinematics are carried out at this stage.
We emphasise this simplification in calculations that involve four leptons. 
With these sets of integrals at hand, we calculate the analytic expressions of master integrals
in terms of generalised polylogarithms up to transcendental weight six~\cite{Goncharov:1998kja}. 
Thus, allowing us to construct the form factors that describe the 
QED processes:
\begin{subequations}
\label{eq:qed_processes}
\begin{align}
\label{eemumu}
e^{+}e^{-}&\rightarrow\mu^{+}\mu^{-}\,,
\\
e^{+}\mu^{-}&\rightarrow e^{+}\mu^{-}\,,
\\
e^{+}e^{-}&\rightarrow e^{+} e^{-}\,,
\\
\label{Compton}
e^{+}e^{-}&\rightarrow\gamma\gamma\,.
\end{align}
\end{subequations}
Lastly, once UV renormalisation and IR subtraction are employed to
construct finite remainders and higher orders in $\epsilon$ for the form factors of the processes~\eqref{eq:qed_processes}, 
we construct their non-vanishing helicity amplitudes.

Due to the similarities between QED and QCD, the results we calculate can, in fact, be derived directly from those presented in \cite{Caola:2021rqz,Caola:2022dfa,Ahmed:2019qtg}. Nonetheless, it is crucial to verify that the same expressions can also be reproduced starting from a purely QED framework. Therefore, we make use of these known results in QCD to validate 
our results. This is carried out by finding linear combinations of Abelian diagrams in QCD.\\

The remainder of the paper is organised as follows.
In Sec.~\ref{techniques}, we discuss the setup of our calculations and the techniques that we utilise to obtain our analytical results.
We outline in 
Sec.~\ref{poles} the procedure to construct the finite remainder and higher orders of $\epsilon$ of the form factors by 
performing UV renormalisation and IR subtraction. 
Then, we discuss our results in Sec.~\ref{results} and perform cross checks against QCD results in Sec.~\ref{checks}.
Lastly, we draw our conclusions and discuss future directions in Sec.~\ref{conclusion}.
Additionally, we include four appendices to provide supplementary details.
In Appendix~\ref{Grouping Results}, we present the grouping of two-loop diagrams involved in processes~\eqref{eemumu} and \eqref{Compton};
Appendix~\ref{Canonical Families} lists the integral families required for these processes;
Appendix~\ref{Coeffs} contains the infrared subtraction constants.
In the arXiv submission of this paper, we include ancillary files containing results for the QED processes discussed in the following sections, with further information on these files provided in Appendix~\ref{Files}.

\section{Formalism and Techniques}\label{techniques}
In this section, we provide an overview of the techniques that we use to perform our calculations. 
We begin by discussing the kinematic setup that we use throughout this paper as well as the perturbative approach to scattering amplitude calculations.
We then discuss the tensor decomposition approach to the calculation of scattering amplitudes that we adopt in this paper.
We then introduce a novel method that allows us to group sub-topologies to parent topologies via shifts in the loop momenta at integrand level using linear algebra.
Lastly, we briefly summarise our adopted strategy 
to reduce two-loop scattering amplitudes to canonical bases of master integrals.
With this formalism at hand, 
we apply these techniques to the processes relevant for our calculation in Sec.~\ref{results}.

\subsection{Kinematics and Setup}

\begin{figure}[t]
    \centering
    \includegraphics[scale=0.17]{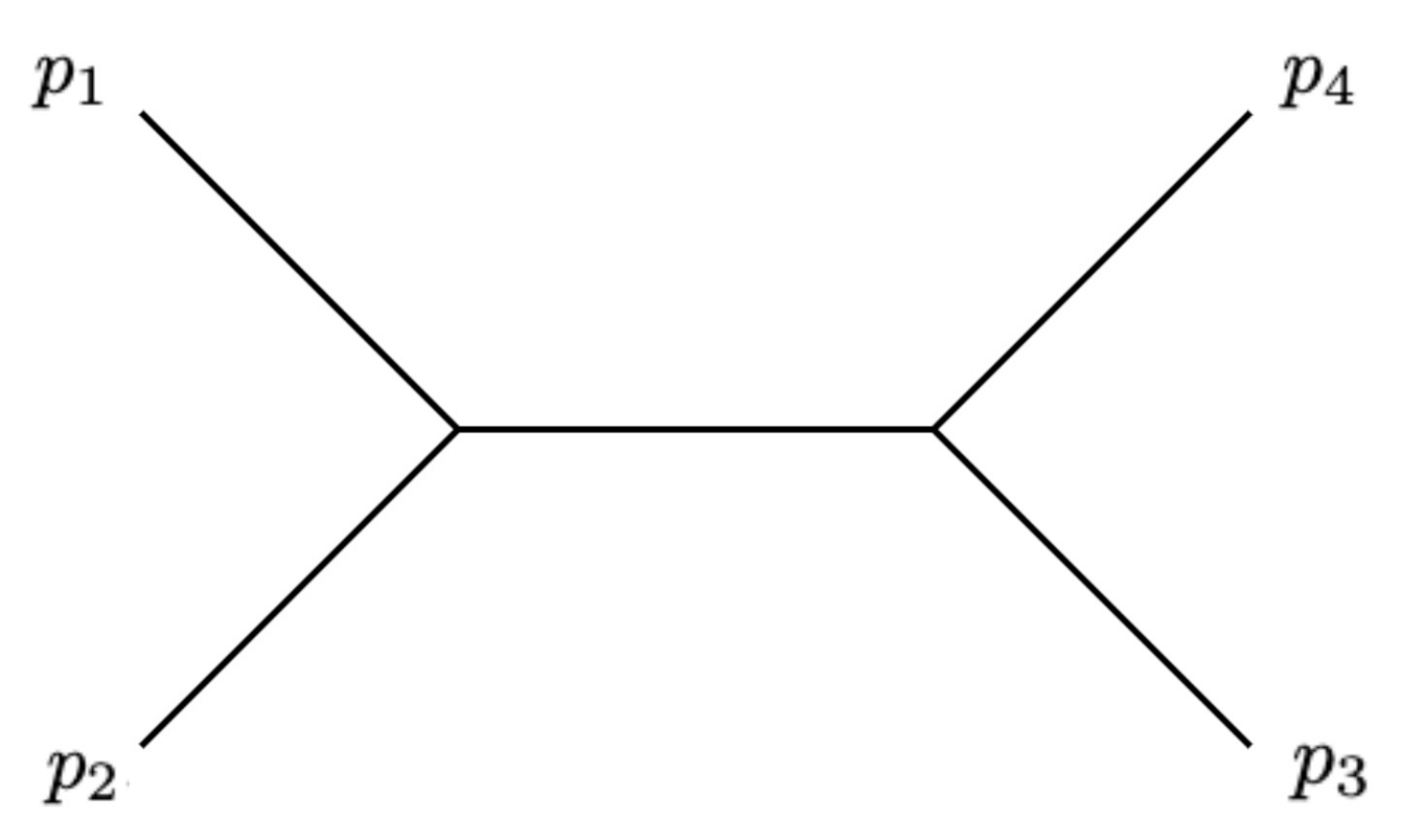}
    \caption{Representative $2\rightarrow 2$ process with the adopted ordering of external momenta. }
    \label{fig:ordering}
\end{figure}

Each of the processes we are considering consists of four particles, we label their corresponding momenta by $p_1$, $p_2$, $p_3$ and $p_4$. The ordering of these momenta can be seen in Fig.~\ref{fig:ordering}.
We consider all external particles to be outgoing so that $p_1=-p_2-p_3-p_4$ is true by momentum conservation. As we are working in massless QED, we have $p_i^2 =0$ for $i=1, 2, 3, 4$.

We also define the Mandelstam variables as:
\begin{subequations}
 \label{eq1}
 \begin{align}
s=(p_3+p_4)^2=2p_3\cdot p_4\,,\label{eq1a}
\\t=(p_2+p_3)^2=2p_2\cdot p_3\,,\label{eq1b}
\\u=(p_2+p_4)^2=2p_2\cdot p_4\,.\label{eq1c}
 \end{align}
\end{subequations}
Similarly, by momentum conservation, 
we can eliminate one of these variables via $u=-s-t$.
Utilising this relation, we are able to express our amplitudes in terms of $s$ and $t$.

We evaluate the processes~\eqref{eq:qed_processes}
in the production region,
\begin{align}
    s>0\,, && t,u<0\,.\label{eq:prod_region}
\end{align}
Because of momentum conservation, 
we introduce a dimensionless variable,
\begin{equation}\label{x_variable}
    x=-\frac{t}{s}\,.
\end{equation}
so that the scattering amplitudes are evaluated in the region $x\in(0,1)$.

Throughout our work, we express scattering amplitudes as an expansion in the bare coupling constant $\alpha_B$,
\begin{equation}
\label{expansion}
    \mathcal{A}_B=\sum^n_{l=0} \left(\frac{\alpha_B}{2\pi}\right)^l\mathcal{A}^{(l)}_B\,,
\end{equation}
where $\mathcal{A}^{(l)}_B$ is the bare amplitude at loop order $l$.

\subsection{Tensor Decomposition}
\label{FFdecomp}

Using symmetries such as Lorentz and gauge invariance we can form a basis of tensor structures ($\mathcal{T}_i$) that comprise the scattering amplitudes we are considering. 
These tensor structures are built by accounting for the  physical states of the ($D$-dimensional) external particles.
We can then break the amplitudes down into form factors, that are completely independent of states of the external particles, which correspond to these tensor structures. We can then construct the scattering amplitude by summing the products of form factors and their corresponding tensor structures,
\begin{equation}\label{FFs}
    \mathcal{A}^{(l)}_B=\sum_{i=1}^{n} \mathcal{F}^{(l)}_{B;i}\, \mathcal{T}_i^{(l)}\,,
\end{equation}
where $\mathcal{F}^{(l)}_{B;i}$ and
$\mathcal{T}^{(l)}_i$
are the form factors and tensor structures 
at a given loop order $l$.

In summary, tensor decomposition allows us to express scattering amplitudes as form factors comprised of scalar products of external and loop momenta. These expressions can be simplified to sums of scalar integrals, whilst encapsulating the complexity of spinors and polarisation vectors within the tensor structures, which we can treat as four-dimensional, and introduce
in the final stages of the calculation. This method is not only beneficial for the purposes of this paper, where we compute helicity amplitudes, but can also be used to compute the interference of scattering amplitudes at loop level.

To obtain these form factors from an amplitude we build a projector operator from the relevant tensor structures. To build this projector operator, we first form a matrix,
\begin{equation}
    M_{ij}^{(l)} = \mathcal{T}^{(l)\dagger}_i\,\mathcal{T}^{(l)}_j\,.
\end{equation}
Then, using this matrix, we obtain the projector operator via,
\begin{equation}
    \mathcal{P}_i^{(l)} = \sum_j \left( M_{ij}^{(l)}\right)^{-1}\mathcal{T}^{(l)}_j\,,
\end{equation}
which, when applied onto the amplitude, 
gives us the form factors,
\begin{equation}
    \mathcal{F}^{(l)}_{B;i} = \mathcal{P}^{(l)}_i\,\mathcal{A}^{(l)}_B\,.
\end{equation}

Referring back to Eq.~(\ref{FFs}), we can see that the only dependence on the configuration of the external states resides within the tensor structures $\mathcal{T}_i$, 
which have until now been considered in $D$ space-time dimensions.
For the purpose of our following calculations, however, 
we are interested in computing physical quantities 
in $D=4$. 
We can benefit from this as we can relate tensor structures with independent helicity configurations of the amplitudes.
This scheme is very similar to the 't Hooft-Veltman scheme, where internal states are considered in $D$ dimensions and external states are considered in four dimensions.

In essence, we perform all of the gamma algebra, Lorentz contractions and compute traces in $D$ dimensions when we are constructing our form factors $\mathcal{F}^{(l)}_{B;i}$ thereby confining all of the $D$-dimensional dependence to them. Whereas we can specify the tensor structures $\mathcal{T}_i$ to be in four dimensions when we fix the helicities of the external states. This corresponds to the dimensional regularisation scheme used in \cite{Peraro:2020sfm}.

This has the benefit of reducing the number of form factors we need to consider. 
Said differently, some tensor structures that are independent in $D$ dimensions become dependent once we restrict them to the four-dimensional space
of external momenta. 
Therefore, these tensor structures are valid at all loop orders and we can safely drop their superscript $(l)$ in 
Eq.~\eqref{FFs}
to only refer to four-dimensional physical processes.

\subsection{Matrix-Based Method for Grouping Feynman Diagrams}\label{groups}

Integration-by-parts (IBP) relations, which will be discussed in Sec.~\ref{IBPs},  are extremely useful in the calculation of scattering amplitudes. However, they can be a bottleneck in many calculations. Hence, it is beneficial to reduce the number of IBP relations we have to find. One way to do this is to group diagrams that obey the same IBP relations. In order to achieve this, we need to find diagrams that share propagators. 
This isn't always immediately obvious as there are multiple ways that we can assign momenta to Feynman diagrams. Therefore, we not only need to consider diagrams that share propagators, but also those that share propagators after a shift in the loop momenta.
This can be a challenging task due to the vast amount of loop momenta shifts that are possible.

For instance, at two loops, 
if we consider a traditional method in which we check every possible shift to see if a match is possible, the number of possible momenta shift permutations that would need to be considered is 59,049.
This would be for a case where we know which ``Parent" diagram a sub-diagram will match to. However, if we consider a case where we don't know which ``Parent" diagram a sub-diagram would map to, then we would need to compare each sub-diagram to many diagrams increasing the permutations we need to consider further.
Therefore, we need an efficient method to find these groupings of diagrams, and the momenta shift required to obtain matching propagators.

We opt for a method to represent the propagators for each diagram as a matrix. Then, through matrix manipulations, we are able to find the possible momenta shifts to obtain matching propagators.

\subsubsection*{Matrix Representation of Feynman diagrams}

The idea is that each row of the matrix represents a propagator. For example,
\begin{align}
(p_2 + p_3 - q_2)^2 
\rightarrow 
\{0,-1,1,1,0\}\,,
\end{align}
where each column in the row corresponds to $q_1$, $q_2$, $p_2$, $p_3$, $p_4$ respectively.

Once we have done this for each propagator we represent 
all propagators in a matrix form,
\begin{gather}
\frac{1}{(q_{1})^{2}(q_{2})^{2}(q_{1}+q_{2})^{2}(p_{4}-q_{2})^{2}(p_{3}+q_{1}+q_{2})^{2}}
\notag\\
\times\frac{1}{(p_{2}+p_{3}+q_{1}+q_{2})^{2}(p_{2}+p_{3}+p_{4}+q_{1})^{2}}
\notag\\
\Downarrow
\notag\\
\begin{pmatrix}-1 & 0 & 0 & 0 & 0\\
0 & -1 & 0 & 0 & 0\\
-1 & -1 & 0 & 0 & 0\\
0 & -1 & 0 & 0 & 1\\
-1 & -1 & 0 & -1 & 0\\
-1 & -1 & -1 & -1 & 0\\
-1 & 0 & -1 & -1 & -1
\end{pmatrix}\label{Example Matrix}
\,,
\end{gather}
where we have ensured that all instances of the loop momenta are negative. In the next section, we discuss how we can use these matrices to find momenta shifts between diagrams.

Although we limit the presentation of our method in this section to the grouping of Feynman diagrams with massless propagators, this representation can be extended to account for massive propagators by simply adding additional columns for the internal masses involved in the process.

\subsubsection*{Finding Momenta Shifts}
The next step is to find the momenta shifts of diagrams with fewer than 7 unique propagators that will make their propagators align with a parent diagram's. To achieve this, we compare the rows of a sub-diagram's matrix, after an arbitrary shift, to that of a parent diagram.

Firstly, let us consider how to represent a shifted matrix. When we shift the loop momenta 
$q_1$ and $q_2$, we end up with,
\begin{align}
q_1 &\rightarrow a q_1 + b q_2 + c p_2 + d p_3 + e p_4\,,
\notag\\
q_2 &\rightarrow f q_1 + g q_2 + h p_2 + i p_3 + j p_4\,,
\end{align}
where $a,b,c,d,e,f,g,h,i,j\in\{-1,0,1\}$. 

In matrix form, this shift becomes, 
\begin{equation}
\begin{pmatrix}
q_1\\
q_2\\
p_2\\
p_3\\
p_4\\
\end{pmatrix} 
\rightarrow 
\begin{pmatrix}
a&b&c&d&e\\
f&g&h&i&j\\
0&0&1&0&0\\
0&0&0&1&0\\
0&0&0&0&1\\

\end{pmatrix} \cdot
\begin{pmatrix}
q_1\\
q_2\\
p_2\\
p_3\\
p_4\\
\end{pmatrix} \,.
\end{equation}
Here and in the following, 
we refer to this matrix as the ``Shift Matrix"\footnote{
This Shift Matrix can also be extended to crossing 
of external kinematics, where additional requirements 
as momentum conservation and on-shell conditions
need to be satisfied.
This is considered with exchanges of external momenta by replacing the identity matrix in the external momentum columns/rows with arbitrary constants.}. 
We then produce a matrix that consists of the coefficients of the shifted loop momenta by multiplying a diagram's matrix representation, for example Eq.~(\ref{Example Matrix}), by the Shift Matrix. 

We can then set up systems of linear equations\footnote{In the two-loop case that we present, we have a system of 25 equations including 10 variables.} by comparing the elements of the shifted diagram with those of a parent diagram. This could appear problematic as a diagram with fewer propagators will have fewer rows to their matrix, so we cannot make a comparison with a parent matrix. To account for this, we make the comparisons with permutations of the rows of the parent matrix. For example, if we want to find the shift of a diagram with 5 propagators to a parent with 7 propagators,  we consider all of the permutations of 5 rows of the parent matrix, and then try to solve the system of linear equations for the variables $\{a, b, c,\hdots, j\}$.
If we find a solution to a system of equations, we have found a possible shift from a sub-diagram to a parent diagram. We opted to solve this system of equations analytically, however numerical methods could also be applied.

This method reduces the amount of permutations that we have to consider, as we are only considering permutations of rows of a matrix, rather than considering every possible expression that each loop momenta could take after a shift.
In the two-loop case, if we consider a traditional method to find the momenta shifts for the two loop momenta, we have 10 variables $\{a, b, c,\hdots, j\}$ that can each take the values $\{-1,0,1\}$. Therefore, we have $3^{10}=59,049$ possible permutations of loop momenta shifts for every sub-diagram. However, using the matrix-based method that we have proposed, we can reduce the permuations that need to be considered to $\frac{7!}{(7-6)!}=5,040$ for diagrams with 6 propagators, $\frac{7!}{(7-5)!}=2,520$ for diagrams with 5 propagators, and $\frac{7!}{(7-4)!}=840$ for diagrams with 4 propagators\footnote{These values are how many permutations have to be considered when checking against one potential parent diagram. To check for shifts against all parent diagrams, we need to multiply these values by the amount of parent diagrams there are for the process.}.

In general, the number of permutations that need to be considered, when comparing against one parent diagram, can be calculated using,
\begin{equation}
\text{Permutations}=\frac{(N_\text{ parent})!}{(N_\text{parent}-N_\text{sub})!}\,,
\end{equation}
where $N_\text{parent}$ and $N_\text{sub}$ correspond to the number of propagators in the parent diagrams and sub-diagrams respectively.

The grouping of diagrams that we find after applying this method to the $e^+e^-\rightarrow\mu^+\mu^-$ and $e^+e^-\rightarrow\gamma\gamma$ processes can be found in Appendix~\ref{Grouping Results}.

This method can be applied to massive cases and higher-loop orders as long as all parameters are accounted for in the shift matrix.
Whilst this is a novel approach, alternative methods to group Feynman diagrams into topological families have been suggested previously by using Symanzik polynomials in \cite{Hoff:2016pot, Gerlach_2023,Shtabovenko:2021hjx,Maierh_fer_2018}.

\subsection{Reduction to Master Integrals}
\subsubsection*{Reduction to scalar integrals}

Once we decompose our amplitudes, that have been generated diagrammatically using {\sc FeynCalc}~\cite{Shtabovenko:2021hjx} and {\sc FeynArts}~\cite{Hahn:2000kx},  into form factors, we obtain expressions that are a sum of integrands. These integrands contain scalar products of the loop momenta $q_i$ and external momenta $p_i$ in the numerator. However, in order compute analytical expressions for the integrands in our form factors it is beneficial to express them as sum of scalar integrals of the form,
\begin{align}
j(&\text{fam},\{a_1,a_2,...,a_N\}) 
\notag\\
&=s^{-l\epsilon}e^{l\epsilon\gamma_E}\prod_{j=1}^{l}\int_{}^{} \left(\frac{\text{d}^Dq_j}{i\pi^{D/2}}\right) \frac{1}{D_1^{a_1}D_2^{a_2}...D_N^{a_N}}\,,
\label{scalar integral}
\end{align}
where ${D_i}$ are the inverse propagators of the internal particles, `$\text{fam}$' is the integral family that the scalar integral belongs to, $l$ denotes the number of loops that appear in the Feynman diagram we are considering, ${a_1,a_2,...,a_N}$ are the powers on the denominators ${D_1,D_2,...,D_N}$ and $\gamma_E$ is the Euler-Mascheroni constant.

To go from our integrands to  scalar integrals, we need to remove all dependence on scalar products of momenta in the numerator. To do this we need to find ways of expressing all of the scalar products in the numerator as linear combinations of the inverse propagators $\{D_i\}$. Depending on the loop order, this can require additional inverse propagators that do not appear in the original integrand expressions. We refer to these additional propagators as auxiliary propagators\footnote{As we have applied momenta shifts to group Feynman diagrams into families, we are able to choose auxiliary propagators that are sufficient for the whole family of diagrams rather than each diagram individiually.}. Take the example of the two-loop case, we will have ${D_1,D_2,\hdots,D_7}$ as the internal propagators of the Feynman diagram, but we require two additional propagators ${D_8, D_9}$ in order to express all scalar products in terms of propagators. Once we have replaced the scalar products with propagators, every term in the expression can then be written as a product of the propagators $\{D_i\}$ to some power as seen in Eq. (\ref{scalar integral}).

Upon expressing our form factors as a sum of scalar integrals, the next required step is to compute these integrals. However, this can become a bottleneck in the calculation. A useful technique to assist in these computations is integration-by-parts reduction.

\subsubsection*{Integration-by-Parts Reduction}\label{IBPs}

Once we have produced expressions for the form factors in terms of scalar integrals, we then begin to compute these integrals. However, given the multitude of scalar integrals that appear in our form factors, this can be quite a cumbersome task.
We utilise the Laporta algorithm \cite{Laporta:2000dsw} to find integration-by-parts relations between integrals, so that we can express a large basis of integrals in terms of a minimal basis of ``Master Integrals" (MIs).
We opt to use the implementation of this algorithm in the {\sc Mathematica} package {\sc LiteRed}~\cite{lee2012presenting} to generate the IBP relations between integrals and MIs. We can then use the package {\sc FiniteFlow} \cite{Peraro_2019} to speedily perform the reduction of form factors to MIs using 
analytical reconstructions over finite fields~\cite{vonManteuffel:2014ixa,Peraro:2016wsq}.

\subsection{Canonical Basis}\label{canonical}

Whilst performing our IBP reduction, we opt to use a canonical basis of master integrals~\cite{Henn:2013pwa}. 
We consider integrals 
that admit a $d\log$ representation~\cite{Henn:2020lye}.
Feynman integrands that admit a $d\log$ can be represented as differential forms, exhibiting a $dx/x$ behaviour in each variable near singularities. Explicitly, 
\begin{equation}
\mathcal{I}^{(l)}=\sum_{k=1}^{l\lfloor D \rfloor } 
c_k \,
d\log\alpha_1^{(k)}\wedge d\log\alpha_2^{(k)}\wedge\hdots\wedge d\log\alpha_n^{(k)}\,,
\end{equation}
with,
\begin{equation}
d= \sum_{i=1}^n dx_i \frac{\partial}{\partial x_i}\,,
\end{equation}
$\alpha_i^{(k)}$ represent the variables we are integrating over, and the wedge is the usual definition of a differential form
giving rise to an oriented volume after integration. The coefficient $c_k$ are the leading singularities of the integrand.

We obtain integrals that admit a $d\log$ representation
using the package {\sc DlogBasis}~\cite{Henn:2020lye}. 
These integrals automatically obey the system of 
differential equations in canonical form, 
\begin{equation}
\label{differential equation}
 \text{d}\textbf{g}(s,t;\epsilon)=\epsilon\, d\tilde{\textbf{A}}\, 
 \cdot \textbf{g}(s,t;\epsilon)\,,
\end{equation}
with, 
\begin{align}
 \tilde{\textbf{A}} &= \sum_{k=1}^{3}
 \textbf{A}_k
 \log\left[ W_k (s,t)\right]\,,
\end{align}
where $\textbf{A}_k$ are $\mathbb{Q}$ matrices, 
$\textbf{g}$ the basis of master integrals,
and $W_k$ represent the letters of the alphabet,
\begin{align}\label{alphabet}
W_1=s\,,&&W_2=t\,, && W_3 = u=-s-t\,.
\end{align}
With the canonical differential equation~\eqref{differential equation}, 
we can express our sets of master integrals as,
\begin{align}
\label{eq:Chenformula}
\textbf{g}\left(s,t;\epsilon\right)
&=\mathbb{P}\exp\left(\epsilon\int_{\mathcal{C}}d\tilde{A}\right)
\textbf{g}_{0}\left(\epsilon\right)\,,
\end{align}
where $\mathbb{P}$ accounts for the path ordering in the matrix exponential along the contour $\mathcal{C}$
in the space of the kinematic variables $s,t$,
and $\textbf{g}_{0}$ represents the boundary values at the base point of the contour $\mathcal{C}$.

By assigning a transcendental weight of $-1$ to $\epsilon$, we can choose a basis $\textbf{g}(s,t;\epsilon)$ that has uniform transcendental weight, such that its expansion around $\epsilon =0$ starts at order $\epsilon^0$,
\begin{equation}
\textbf{g}(s,t;\epsilon) =\sum_{w\geq 0} \epsilon^{w} \textbf{g}^{(w)}(s,t)\,,
\end{equation}
where $\textbf{g}^{(w)}(s,t)$ are the terms in the expansion of $\textbf{g}(s,t;\epsilon)$ at transcendental weight $w$. 
For the purpose of the calculations in this work, 
we integrate our MIs up to transcendental weight six.

At one-loop, poles start at $\mathcal{O}\left(\epsilon^{-2}\right)$ and at two-loop poles start at $\mathcal{O}\left(\epsilon^{-4}\right)$. Expanding up to weight 6 gives expressions for the one-loop integrals up to $\epsilon^4$ and two-loop integrals up to $\epsilon^2$.

In order to solve the differential equations~(\ref{differential equation}), we need to provide boundary conditions at each order in $\epsilon$.
For the sake of the simplicity and to systematically 
account for analytic continuation of integrals, 
we use the package {\sc AMFlow}~\cite{Liu:2017jxz,Liu_2023} to evaluate our MIs phase-space points in the production region~\eqref{eq:prod_region}.

We express the analytical expressions for our MIs in terms of generalised polylogarithms (GPLs)~\cite{Goncharov:1998kja}
as function of the dimensionless variable $x$
of Eq.~\eqref{x_variable}.
GPLs can be expressed as the recursively iterated integrals,
\begin{equation}
G(a_1,\hdots,a_n;x)=\int_0^x \frac{dt}{t-a_1} G(a_2,\hdots,a_n;t)\,,
\end{equation}
where at least one $a_i\neq 0$. We also define $G(;t)=1$.
The length of the vector $\vec{a} =(a_1,a_2,...,a_n)$ defines the transcendental weight of the GPL.
In the case where all $a_i=0$, we define,
\begin{equation}
G(0_1,0_2,\hdots,0_n;x)=\frac{1}{n!} \log^n(x)\,,
\end{equation}
where $n$ is the number of $0$s.

We construct a minimal basis of GPLs up to transcendental weight six functions, 
by systematically enumerating Lyndon words on the sets 
$\{0,1\}$ through the function 
\verb"DecomposeToLyndonWords"
of {\sc PolyLogTools}~\cite{Duhr_2019}.
This means that we are able to express all form factors in terms of at most 23 GPLs, which makes numerical evaluation of our final results more efficient.

\subsubsection*{One-Loop Canonical Basis}
\begin{figure}[t]
\center
\includegraphics[scale=0.25]{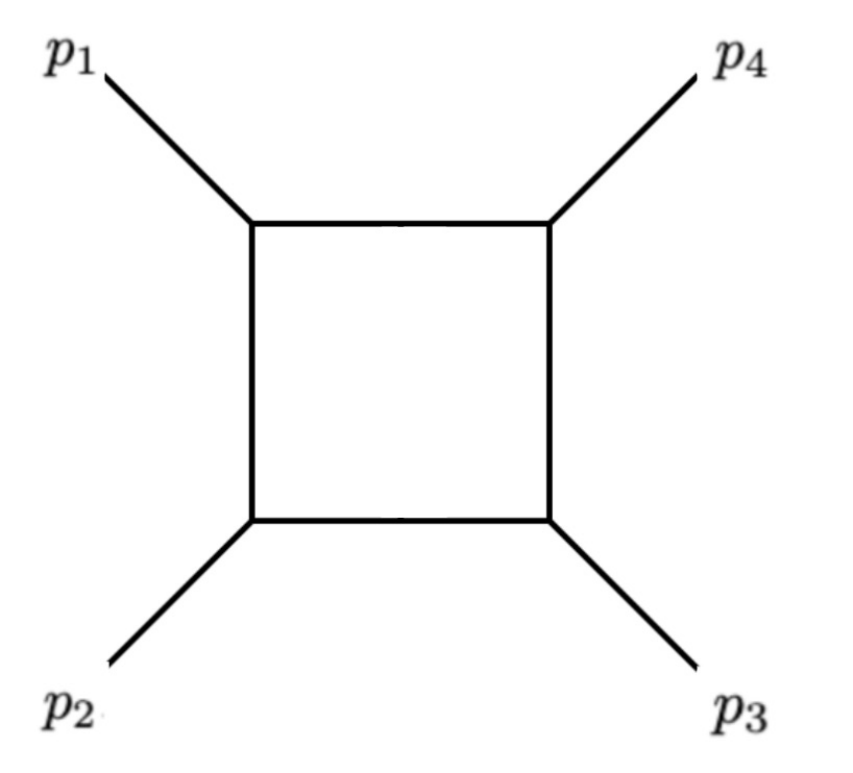}
\caption{One-loop box with the ordering of external momenta of Fig.~\ref{fig:ordering}.}\label{fig:OneLoop}
\end{figure}

We can express all one-loop Feynman integrals within the scattering amplitudes in terms of 5 MIs. These integrals stem from the one-loop box shown in Fig. \ref{fig:OneLoop} and the diagram that results from exchanging the external legs with momenta $p_3$ and $p_4$, which we will henceforth refer to as the ``crossed box".
The complete set of propagators that are used to perform IBP reduction on these diagrams are seen in
Table~\ref{OneLoopProps}.
\begin{table}[t]
\center\begin{tabular}{|l||l|}
\hline
\textbf{Denominator} & \textbf{Propagator}            \\ \hline\hline
$D_1$            & $(q)^2$                        \\ 
$D_2$            & $(q_1-p_2)^2$ \\ 
$D_3$            & $(q-p_2-p_3)^2$        \\ 
$D_4$            & $(q-p_2-p_3-p_4)^2$                        \\ \hline
\end{tabular}
\caption{Propagators for the one-loop box of Fig.~\ref{fig:OneLoop}.}\label{OneLoopProps}
\end{table}
From this propagator configuration, we obtain 3 MIs which we shall denote 
as $d\log(\text{1234},i)$, with $i= 1, 2, 3$. These MIs correspond respectively to the one-loop box in the $(s,t)$-channel, the one-loop triangle in the $t$-channel and the one-loop triangle in the $s$-channel.

To find the MIs that are obtained from the crossed box we can simply perform an exchange of $p_3$ and $p_4$ in the propagators, as well as performing an exchange of $t$ and $u=-s-t$ in any kinematic pre-factors to the MIs produced by the one-loop box. These integrals will be denoted as $d\log(\text{1243},i)$, where $i= 1, 2, 3$.

Altogether we have 6 canonical integrals, but due to Lorentz symmetries $d\log(\text{1234},3)$ is identical to $d\log(\text{1243},3)$ when $p_3$ and $p_4$ are exchanged. Therefore we are left with 5 canonical MIs.

\subsubsection*{Two-Loop Canonical Basis}
At two-loop level, all integrals contained with the scattering amplitudes of both the $e^+e^-\rightarrow \mu^+ \mu^-$ and $e^+e^-\rightarrow \gamma \gamma$ processes can be expressed in terms of MIs that are derived from two diagrams and permutations of their external momenta. This allows us to split the MIs into two cases, those that come from planar (PL) diagrams and those that come from non-planar (NPL) diagrams.

\begin{figure}[t]
    \centering
    \begin{subfigure}[b]{0.22\textwidth}
        \centering
        \includegraphics[scale=0.5]{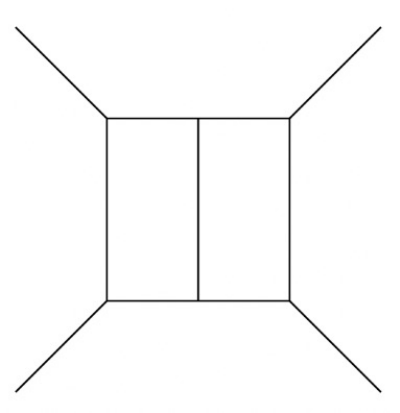}
    \end{subfigure}
    \hfill
    \begin{subfigure}[b]{0.25\textwidth}
        \centering
        \includegraphics[scale=0.5]{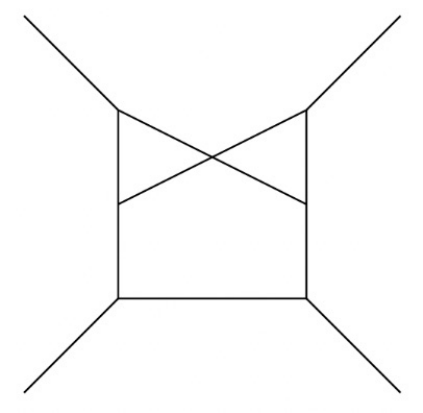}
    \end{subfigure}
    \caption{The two-loop topologies that define our planar (PL) and non-planar (NPL) integral families.}
    \label{fig:PLandNPL}
\end{figure}

Let us consider the planar and non-planar diagrams that have the external momenta configuration of Fig.~\ref{fig:ordering} that can be seen in Fig.~\ref{fig:PLandNPL}.
The complete set of propagators ${D_i}$ that are used to perform the IBP reduction for these diagrams are summarised in Table~\ref{Props}.
\begin{table}[t]
\center\begin{tabular}{|l||l||l|}
\hline
\textbf{Diagram} & \textbf{PL}       & \textbf{NPL}      \\ \hline\hline
$D_1$            & $(q_1)^2$             & $(q_1)^2$             \\ 
$D_2$            & $(q_1-p_2-p_3-p_4)^2$ & $(q_1-p_2-p_3-p_4)^2$ \\ 
$D_3$            & $(q_1-p_3-p_4)^2$      & $(p_4-q_1+q_2)^2$     \\ 
$D_4$            & $(q_2)^2$             & $(q_2)^2$             \\ 
$D_5$            & $(q_2-p_3-p_4)^2$     & $(q_2-p_2-p_3)^2$     \\ 
$D_6$            & $(q_2-p_4)^2$         & $(q_2-p_3)^2$         \\ 
$D_7$            & $(q_1-q_2)^2$         & $(-q_1+q_2)^2$        \\ 
$D_8$            & $(q_1-p_4)^2$         & $(q_1-p_3)^2$         \\ 
$D_9$            & $(q_2-p_2-p_3-p_4)^2$ & $(q_2-p_2-p_3-p_4)^2$ \\ \hline
\end{tabular}
\caption{Propagators for planar and non-planar diagrams in the configuration of external momenta of Fig.~\ref{fig:ordering}.}
\label{Props}
\end{table}

The planar diagram shown above provides 8 MIs that we denote 
by $d\log(\text{PL1234},i)$, where $i= 1, 2, 3,\hdots,8$.
Similarly, the non-planar diagram shown above provides 12 MIs that we denote 
by $d\log(\text{NPL1234},i)$, where $i= 1, 2, 3,\hdots,12$.

If we are to consider just these two diagrams, we will not have a sufficient basis to express all integrals in terms of MIs. Therefore, we must also consider the permutations of the external momenta. If we fix $p_1$ in its position and permute the remaining external momenta we create six families each for both the planar and the non-planar cases.

\begin{figure}[t]
\center\includegraphics[scale=0.22]{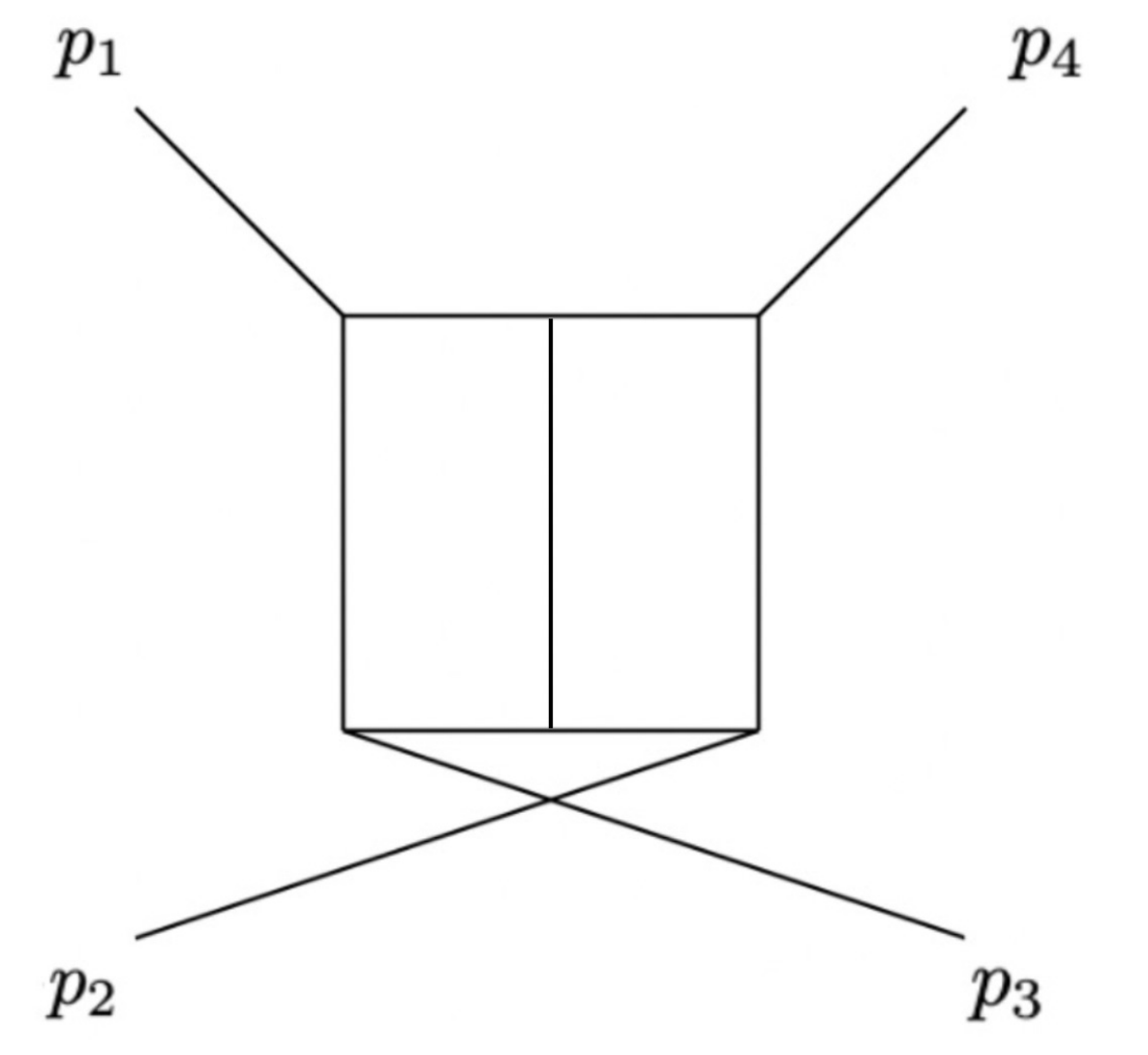}
\caption{Two-loop planar diagram with an exchange of $p_2$ and $p_3$ that defines our integral family PL1324.
}\label{fig:PL32}
\end{figure}

Let us consider the case where the external momenta configuration in which $p_2$ and $p_3$ have been exchanged, as seen in Fig.~\ref{fig:PL32}.
This diagram will produce an additional 8 canonical integrals. It is possible to derive these MIs from the basis 
 $d\log(\text{PL1234},i)$ by simply exchanging $p_2$ with $p_3$ in each of the propagators shown in Table~\ref{Props} as well as performing the exchange of $s$ and $u=-s-t$ in the kinematical pre-factors. We denote these MIs as $d\log(\text{PL1324},i)$, with $i= 1, 2, 3,\hdots,8$.

Repeating this process for all possible permutations of both the planar and non-planar case gives us 120 integrals. We then look to see if any of these integrals have symmetries or relations to each other. We prioritise planar integrals over non-planar integrals.
Once we have considered relations and symmetries between these integrals we are able to express all integrals in terms of a minimal set of at most 39 MIs. Further details on the MIs for the two-loop $e^+e^-\rightarrow\mu^+\mu^-$ and $e^+e^-\rightarrow\gamma\gamma$ processes can be found in Appendix~\ref{Canonical Families}. 

\section{UV renormalisation and IR subtraction}\label{poles}

Once we have expressed our form factors as an expansion in the bare coupling constant~(\ref{expansion}), we obtain expressions that contain $\epsilon$ poles in the dimensional regulator.\footnote{We can also apply the same renormalisation techniques directly on bare scattering amplitudes $\mathcal{A}_B$.}
In this section we discuss the techniques we use to remove these poles.

The singularities in our expressions appear in two forms: UV and IR divergences.
UV divergences are due to divergences in the integrals when loop momenta become infinite, whereas IR divergences stem from the soft emissions and collinear singularities. We will first address how we remove the UV poles, followed by the  IR poles.

To maintain a general perspective in this section, we will refer to scattering amplitudes. This approach is valid since both UV renormalization and IR subtraction can be applied to scattering amplitudes and form factors in a similar manner.

\subsection{UV Renormalisation}

We absorb UV divergences by renormalising the couple constant in the $\overline{\text{MS}}$ scheme. In detail, we exchange 
bare  with a renormalised coupling constant as follows,
\begin{align}
   & \alpha_B \mu_0^{2\epsilon}S_\epsilon = 
   \alpha\,\mu^{2\epsilon}
    \\
   & \times\left[ 1-\frac{\beta_0}{\epsilon}\left(\frac{\alpha}{2\pi}\right)+\left(\frac{\beta_0^2}{\epsilon^2}-\frac{\beta_1}{2\epsilon}\right)\left(\frac{\alpha}{2\pi}\right)^2 +\mathcal{O}\left ( \alpha^3\right) \right]\,,
   \nonumber
\end{align}
where  $\alpha$ is the renormalised coupling constant, $\mu$ is the renormalisation scale, $S_\epsilon=\exp(-\epsilon\gamma_E)$ is the normalisation factor. Additionally, we have the $\beta$ functions, 
\begin{equation}
\beta_0 = -\frac{2N_f}{3} \text{ and } \beta_1 = -N_f\,,
\end{equation}
with $N_f$ the number of fermions.

After performing the exchange and re-writing our amplitudes as an expansion in $\left(\frac{\alpha}{2\pi}\right)$, we can find the renormalised $l$-loop amplitudes by considering the coefficient of $\left(\frac{\alpha}{2\pi}\right)^{(l)}$.
From this we obtain the following relations between the bare amplitudes and the renormalised amplitudes,
\begin{subequations}
\begin{align}
&\mathcal{A}^{(0)} = \mathcal{A}_B^{(0)}\,,&&\\
&\mathcal{A}^{(1)} = \mathcal{A}_B^{(1)}-\frac{\beta_0}{\epsilon}\mathcal{A}_B^{(0)}\,,&&\\
 &\mathcal{A}^{(2)} = \mathcal{A}_B^{(2)}-\frac{2\beta_0}{\epsilon}\mathcal{A}_B^{(1)}+\left(\frac{\beta_0^2}{\epsilon^2}-\frac{\beta_1}{2\epsilon}\right)\mathcal{A}_B^{(0)}\,,
\end{align}
\label{eq:uv_renormalisation}
\end{subequations}
where $\mathcal{A}^{(l)}$ are the $l$-loop renormalised amplitudes.

When we substitute the expressions for our bare amplitudes into these equations, we obtain the renormalised amplitudes that contain no UV poles.
\subsection{IR Subtraction}
Once we have expressions for our renormalised amplitudes, we can now remove the remaining poles using IR subtraction in the Soft-Collinear Effective Theory (SCET) formalism \cite{Becher_2015}.

We should note that we have utilised techniques that are relevant in QCD to obtain many of the expressions required to remove the remaining poles. We then abelianise these expressions to get the relevant expressions for our QED processes. Therefore, our derivations may include terminology relevant in QCD rather than QED.

The finite remainder of an amplitude can be written as,
\begin{equation}
    \mathcal{R}(\{p\},\mu)=\lim_{\epsilon \to 0} \mathcal{Z}^{-1}(\epsilon,\{p\},\mu)\, \mathcal{A}(\epsilon,\{p\})\,,
\end{equation}
where $\mathcal{Z}$ is the multiplicative colour-space operator.

We can apply this at each loop order to obtain the following relations:
\begin{subequations}
\begin{align}
\mathcal{R}^{(0)} &=\mathcal{A}^{(0)}\,, \\
\mathcal{R}^{(1)} &=\mathcal{A}^{(1)}-\mathcal{I}^{(1)}\mathcal{A}^{(0)}\,, \\
 \mathcal{R}^{(2)} &=\mathcal{A}^{(2)}-\mathcal{I}^{(1)}
 \mathcal{A}^{(1)}-\mathcal{I}^{(2)}\mathcal{A}^{(0)}\,,
\end{align}
\label{Rs}
\end{subequations}
where $\mathcal{I}^{(n)}$ are the subtraction operators, 
\begin{subequations}
\begin{align}
\mathcal{I}^{(1)} &=\mathcal{Z}^{(1)}\,, \\
\mathcal{I}^{(2)} &=\mathcal{Z}^{(2)}-\left( \mathcal{Z}^{(1)}\right)^2\,.
\end{align}
\label{Is}
\end{subequations}
The expressions for $\mathcal{Z}^{(m)}$ operators themselves can be found by solving the equation,
\begin{equation}
\mathcal{Z}(\epsilon,\{p\},\mu)= \mathbb{P}\,\text{exp}\left[\int_{\mu}^{\infty}\frac{d\mu^{'}}{\mu}\Gamma(\{p\},\mu)\right]\,.
\end{equation}
In this equation, the anomalous dimension matrix $\Gamma$ is defined as,
\begin{equation}
\Gamma=\Gamma_{\text{dipole}}(\{p\},\mu) +\Delta_4\,,
\end{equation}
where $\Gamma_{\text{dipole}}$ corresponds to the colour dipole correlations and $\Delta_4$ contains the quadrupole colour correlations.

Once we have abelianised our expressions, $\Delta_4$ vanishes. Therefore, we only need to consider $\Gamma_{\text{dipole}}$ in this work. We will refer to $\Gamma_{\text{dipole}}$ as $\Gamma$ from this point onwards.

The explicit expression for the anomalous dimension $\Gamma$ reads as,
\begin{align}
\Gamma(\{p\},\mu) & =\sum_{1\leq i<j\leq n}T_{i}^{a}T_{j}^{a}\,\gamma^{\text{cusp}}\,L_{ij}+\sum_{i}^{n}\gamma^{i}\,,\label{Gamma}
\end{align}
with, 
\begin{align}
L_{ij} & =\log\left(\frac{\mu^{2}}{-s_{ij}-i0}\right)\,,
\end{align}
$s_{ij}$ are Mandelstam invariants,
\begin{equation}
s_{12}=s_{34}=s\text{ , }s_{23}=s_{14}=t\text{ , and }s_{13}=s_{24}=u\,,
\end{equation}
and $T^{a}_i$ are the generators of the $\mathsf{SU}(N_c)$
 group, 
associated with the particle $i$.
The second term in Eq.~\eqref{Gamma} is dependent on the external particles in the process. Therefore, $\Gamma$ differs when we consider different processes.
These generators are written in terms of the Casimir invariants $C_A$ and $C_F$.
Because we are focused on QED processes, 
we abelianise by setting $C_A\to 0$ and $C_F\to 1$, 
and $T_i\to q_i$, corresponding to the charge of the $i$-th external particle.

In Eq.~\eqref{Gamma}, $\gamma^{\text{cusp}}$ is the cusp anomalous dimension coefficient and $\gamma^{i}$ is the anomalous dimension coefficient for particle $i$.

Additionally, the IR renormalisation factor $Z_\text{IR}$ is
defined by the exponentiation of, 
\begin{align}
\log Z_\text{IR} &= \sum_{k=1}^{l}
\left(\frac{\alpha}{2\pi}\right)^l\mathcal{Z}^{(l)}\,.
\label{eq:ZIR}
\end{align}
with
\begin{subequations}
\begin{align}
 &\mathcal{Z}^{(1)}= \frac{\Gamma_0^{'}}{4\epsilon^2}+\frac{\Gamma_0}{2\epsilon}\,, &&\\
&\notag\mathcal{Z}^{(2)}=  \frac{{\Gamma_0^{'}}^2}{32\epsilon^4}+\frac{\Gamma_0^{'}}{8\epsilon^3}\left(\Gamma_0-\frac{3}{2}\beta_0 \right)&&\\
&\hspace{1.5cm}+\frac{1}{4\epsilon^2}\left(-\beta_0\Gamma_0+\frac{\Gamma_0^2}{2}+\frac{\Gamma_1^{'}}{4} \right)+\frac{\Gamma_1}{4\epsilon}\,.&&
\end{align}
\end{subequations}
By substituting the corresponding anomalous dimension coefficients for the relevant process, which can be found in  Appendix \ref{Coeffs}, and performing the IR subtraction, we are able to remove all remaining divergences and obtain expressions for our processes that are of $\mathcal{O}(\epsilon^0)$ or higher.
This cancellation of IR poles serves as a consistency check, as it confirms that the structure predicted by lower loop orders is indeed reproduced through direct calculation.

\section{Results}\label{results}
In this section, we shall use our results from the computation of form factors to present the helicity amplitudes, for non-vanishing helicity configurations, for all processes~\eqref{eq:qed_processes}.
Additionally, we present the bare one-loop form factors, in terms of integrals, for the $e^+e^-\rightarrow \mu^+ \mu^-$ and the $e^+\mu^-\rightarrow e^+ \mu^-$ scatterings to illustrate how we can exchange momenta to obtain results for different processes. Expressions for the one- and two-loop form factors for the $e^+e^-\rightarrow \mu^+ \mu^-$, $e^+\mu^-\rightarrow e^+ \mu^-$ and $e^+e^- \rightarrow \gamma\gamma$ processes can be found in the ancillary files. We do not provide the bare form factors for $e^+e^-\rightarrow e^+e^-$ as we obtain the helicity amplitudes for this process from the form factors of $e^+e^-\rightarrow \mu^+ \mu^-$ and $e^+\mu^-\rightarrow e^+ \mu^-$ after UV renormalisation and IR subtraction.

At this stage we opt to introduce the dimensionless variable $x$~\eqref{x_variable}, which roughly corresponds to setting 
$s=1$, and allows to reduce the alphabet~(\ref{alphabet}) to two letters, $\{x,1-x\}$. The dependence on $s$ is easily recovered by dimensional analysis. 

Furthermore, we introduce an additional subscript to the form factors $\mathcal{F}_i$ to help us distinguish between the processes:\footnote{These form factors have undergone UV renormalisation and IR subtraction, thus will contain no poles.}

\begin{itemize}
\item 
$\mathcal{F}_{\mu\mu,i}$ for $e^+e^-\rightarrow \mu^+\mu^-$. 
\item 
$\mathcal{F}_{e\mu,i}$ for $e^+\mu^-\rightarrow e^+\mu^-$.
\item 
$\mathcal{F}_{\gamma\gamma,i}$ for $e^+e^-\rightarrow \gamma\gamma$. 
\end{itemize}
We introduce a similar notation for the tensor structures $\mathcal{T}_i$. 

We find that for processes involving four fermions we obtain expressions for the form factors and helicity amplitudes that take the form:
\begin{align}\label{Form4fermion}
\notag &\text{Tree-Level}\sim A^{(0)}_{f_3 f_4}\,,&&\\
\notag &\text{One-Loop}\sim A^{(1)}_{f_3 f_4}+N_f\, B^{(1)}_{f_3 f_4}\,,&&\\
&\text{Two-Loop}\sim A^{(2)}_{f_3 f_4}+N_f\, B^{(2)}_{f_3 f_4}+
N_f^2\,C^{(2)}_{f_3 f_4}\,,&&
\end{align}
where $f_i$ is the flavour of the fermion with momenta $p_i$.

Similarly, for $e^+e^- \rightarrow \gamma \gamma$ we find that the form factors and helicity amplitudes take the form:
\begin{align}\label{FormCompton}
\notag &\text{Tree-Level}\sim A^{(0)}_{\gamma\gamma}\,,&&\\
\notag &\text{One-Loop}\sim A^{(1)}_{\gamma\gamma}\,,&&\\
&\text{Two-Loop}\sim A^{(2)}_{\gamma\gamma}+N_f\, B^{(2)}_{\gamma\gamma}\,.
\end{align}

\subsection{$e^+e^-\rightarrow \mu^+\mu^-$}

For the $e^+e^-\rightarrow \mu^+\mu^-$ scattering there are 2 independent Lorentz structures that can appear when we consider the external momenta to be in four space-time dimensions,
\begin{align}
\notag &T_{\mu\mu,1}=\overline{u}(p_2)\gamma_{\mu} u(p_1)\times\overline{u}(p_4)\gamma^{\mu} u(p_3)\,,&&\\
&
T_{\mu\mu,2}=\overline{u}(p_2)\fsl{p_3} u(p_1)\times\overline{u}(p_4)\fsl{p_1} u(p_3)\,.&&
\end{align}
Using the techniques outlined in Sec. \ref{FFdecomp}, we can construct projectors to obtain the tensor decomposition for this process.

For these tensor structures there are 4 non-vanishing helicity configurations:
\begin{equation}
    \left\{-++-,-+-+,+--+,+-+-\right\}\,.
\end{equation}
In our results, we shall present the helicity amplitudes of the first two configurations, as the latter two can be obtained by applying parity to these helicity configurations. 
We achieve this by taking the complex conjugate of these configurations.

To form an expression for the helicity amplitude from the form factors, we need to find a consistent representation for the tensor structures. Using spinor helicity formalism \cite{Dixon:1996wi}, we can express both tensor structures in the form of:
\begin{equation}\label{eemumuTS}
\left\{\Phi_{\mu\mu,(-++-)},\Phi_{\mu\mu,(-+-+)}\right\}=\left\{\left\langle 14 \right\rangle\left[ 23\right]  
,\left\langle 13 \right\rangle\left[ 24\right]\right\}\,.
\end{equation}

Expressing the tensor structures in these forms introduces pre-factors in terms of the Mandelstam variables to our form factors. Specifically, to obtain the helicity amplitudes for our two configurations, we need to sum our form factors as,
\begin{align}
\notag &\mathcal{H}^{(l)}_{\mu\mu,(-++-)}= -2 \mathcal{F}^{(l)}_{\mu\mu,1} +\left(1-x\right)\mathcal{F}^{(l)}_{\mu\mu,2}\,,&&\\&
\mathcal{H}^{(l)}_{\mu\mu,(-+-+)}= -2 \mathcal{F}^{(l)}_{\mu\mu,1} -x\mathcal{F}^{(l)}_{\mu\mu,2}\,.&&
\label{eq:FF_mumu}
\end{align}
We then multiply these expressions by their corresponding spinor structures,
\begin{align}
&\mathcal{A}^{(l)}_{\mu\mu,(\lambda_1\lambda_2\lambda_3\lambda_4)}= \mathcal{H}^{(l)}_{\mu\mu,(\lambda_1\lambda_2\lambda_3\lambda_4)}\Phi_{\mu\mu,(\lambda_1\lambda_2\lambda_3\lambda_4)}\,,&&
\end{align}
where $\lambda_i$ corresponds to the helicity of the $i$-th external particle with momentum $p_i$.
These expressions for the helicity amplitudes, in terms of form factors, are valid at all loop orders.

\newpage

\subsubsection*{One-Loop bare form factors}

Here we present the one-loop bare form factors, $\mathcal{F}_{B;\mu\mu,i}^{(1)}$,\footnote{We have added the subscript $B$ to clarify that these are expressions for the bare form factors.} in terms of master integrals, 
\begin{widetext}
\begin{align}\label{F_eemumu1}
\notag &\mathcal{F}_{B;\mu\mu,1}^{(1)}=\frac{1}{2\epsilon^2 (1-2\epsilon)s}
\left[d\log(1234,1)\left(\frac{\epsilon s -2s -4\epsilon u}{2u}\right)+d\log(1234,2)\left(\frac{2s-\epsilon s -2u(\epsilon -2)}{u}\right)\right.&&\\ \notag\\
&\hspace{2cm}\left.+d\log(1234,3)\left(\xi\left( (2\epsilon^2-\epsilon+2)+\frac{ 2\epsilon(\epsilon-1)}{(2\epsilon-3)}N_f\right) -\frac{(\epsilon-2)s}{u}\right)\right] - \left\{p_3\leftrightarrow p_4,\text{ } \xi \rightarrow -\xi\right\}\,, 
&&
\end{align}
we then must set $\xi=1$.

\begin{align}\label{F_eemumu2}
\notag&\mathcal{F}_{B;\mu\mu,2}^{(1)}=\frac{1}{\epsilon^2 (1-2\epsilon)tu}\left[-d\log(1234,1)\left(\frac{ \epsilon^2 s^2-\epsilon s t-4 \epsilon t^2+2 s t+4 t^2}{2 s u}\right)\right.&&\\ \notag\\
&\left.+d\log(1234,2)\frac{\left(2 \epsilon^2 s^2+\epsilon^2 s t+\epsilon s t-2 \epsilon t^2+2 s t+4 t^2\right)}{s u}+d\log(1234,3)\left( (4-6\epsilon+\epsilon^2)+\frac{(\epsilon-2)(\epsilon-1)s}{u}\right)\right]
\notag
&&\\
&+\left\{p_3\leftrightarrow p_4\right\} \,. &&
\end{align}
\end{widetext}
The exchange of $p_3$ and $p_4$ in both form factors implies the exchange of $t$ and $u$.

Let us first discuss the integrals that are part of the family 1234,
\begin{align}
  \notag & d\log(1234,1)= \epsilon^2 st\, j(1234,\{1,1,1,1\})\,,&&\\
  \notag & d\log(1234,2)= \epsilon^2 t\, j(1234,\{1,1,1,0\})\,,&&\\
 &d\log(1234,3) = \epsilon^2 s\, j(1234,\{1,1,0,1\})\,.\label{fam1234}
\end{align}
The denominators for this family of integrals can be found in Table~\ref{OneLoopProps}.
We are able to obtain the integrals that are part of the family 1243 as follows:
\begin{align}
d\log(1243,i) = d\log(1234,i)\Big|_{p_3\leftrightarrow p_4}
\,,
\label{eq:will_38}
\end{align}
for $i=1,2,3$.
The denominators for this family of integrals can be found by exchanging $p_3$ and $p_4$ in the denominators presented in Table~\ref{OneLoopProps}.

\subsubsection*{Two-Loop bare form factors}
The two-loop bare form factors for $e^+e^-\rightarrow\mu^+\mu^-$ are provided in the ancillary files. These form factors consist of the two-loop integrals discussed in Sec.~\ref{canonical}.

Similarly to the one-loop case, we are able to obtain the MIs for two of these integral families: PL1234 and NPL1234. Then we can perform swaps of external momenta to obtain the master integrals for the remaining families. For example:
\begin{align}
d\log(\text{PL1243},i) = d\log(\text{PL1234},i)\Big|_{p_3\leftrightarrow p_4}
\,,
\end{align}
for $i=1,2,3,\hdots,8$. Also, 
\begin{align}
d\log(\text{NPL1243},i) = d\log(\text{NPL1234},i)\Big|_{p_3\leftrightarrow p_4}
\,,
\end{align}
for $i=1,2,3,\hdots,12$.

Additional information about the two-loop integral families required for the $e^+ e^-\rightarrow \mu^+\mu^-$ can be found in Appendix~\ref{Canonical Families}.

\subsection{$e^+\mu^-\rightarrow e^+\mu^-$}
As we are considering the massless case for each process, the $e^+\mu^-\rightarrow e^+\mu^-$ scattering amplitude may be derived directly from the $e^+e^-\rightarrow \mu^+\mu^-$ as both processes consider a four fermion $2\rightarrow2$ scattering.

Compared to the $e^+e^- \rightarrow \mu^+\mu^-$ scattering, the flavours of the particles with momenta $p_2$ and $p_4$ are swapped. Therefore, to obtain the form factors at all loop levels for the $e^+\mu^- \rightarrow e^+\mu^-$ we can simply exchange the momenta $p_2$ and $p_4$. This exchange leads to several changes to our expressions. Firstly, we must swap the Mandelstam variables $s$ and $t$. Secondly, after swapping the external momenta, the propagators in the integrals will also change. Therefore, we must perform an exchange of $p_2$ and $p_4$ on the integral families in our expressions before computing them. 

There are 4 non-vanishing helicity configurations for this process,
\begin{equation}
    \left\{--++,-+-+,++--,+-+-\right\}\,.
\end{equation}
Similarly to the $e^+e^-\rightarrow\mu^+\mu^-$ process, in our results we present helicity amplitudes for the first two configurations,
\begin{equation}
\left\{\Phi_{e\mu,(--++)},\Phi_{e\mu,(-+-+)}\right\}=\left\{\left\langle 12 \right\rangle\left[ 34\right] \text{, } \left\langle 13 \right\rangle\left[ 24\right]\right\}\,.
\end{equation}
To obtain the corresponding helicity amplitudes for these configurations we need to sum the form factors with the following pre-factors:
\begin{align}
\notag &\mathcal{H}^{(l)}_{e\mu,(--++)}=2 \mathcal{F}^{(l)}_{e\mu,1} -\left(1-x\right)\mathcal{F}^{(l)}_{e\mu,2} \,,&&
\\&\mathcal{H}^{(l)}_{e\mu,(-+-+)}=2 \mathcal{F}^{(l)}_{e\mu,1} -\mathcal{F}^{(l)}_{e\mu,2} \,.&&
\end{align}
We then multiply these expressions by their corresponding spinor structures,
\begin{align}
&\mathcal{A}^{(l)}_{e\mu,(\lambda_1\lambda_2\lambda_3\lambda_4)}= \mathcal{H}^{(l)}_{e\mu,(\lambda_1\lambda_2\lambda_3\lambda_4)}\Phi_{e\mu,(\lambda_1\lambda_2\lambda_3\lambda_4)}\,.&&
\end{align}

\subsubsection*{One-Loop bare form factors}

Here we present the one-loop bare form factors, $\mathcal{F}_{B;e\mu,i}^{(1)}$, 
in terms of integrals.
These form factors can be directly obtained from 
$\mathcal{F}_{B;\mu\mu,i}^{(l)}$ as follows, 
\begin{align}
\mathcal{F}_{B;e\mu,1}^{(l)} &= \mathcal{F}_{B;\mu\mu,1}^{(l)}\Big|_{p_2\leftrightarrow p_4}\,,\\
\mathcal{F}_{B;e\mu,2}^{(l)} &= \mathcal{F}_{B;\mu\mu,2}^{(l)}\Big|_{p_2\leftrightarrow p_4}\,,
\end{align}
where the exchanging of $p_2$ and $p_4$ is understood as the exchange of $s$ and $t$. 

At one-loop, the form factors $F_{B;e\mu,i}^{(1)}$ become, 
\begin{widetext}
\begin{align}
 & \mathcal{F}_{B;e\mu,1}^{(1)}=\frac{1}{2\epsilon^{2}(1-2\epsilon)t}\left[d\log(1432,3)\left(\xi\left((2-\epsilon+2\epsilon^{2})+\frac{2\epsilon(\epsilon-1)}{2\epsilon-3}N_{f}\right)-\frac{(\epsilon-2)t}{u}\right)\right.\label{F_emuemu1}\\
\nonumber \\
 & \hspace{2cm}\left.+\frac{1}{2}d\log(1432,1)\left((2-5\epsilon)-\frac{(\epsilon-2)s}{u}\right)+d\log(1432,2)\left(2-\epsilon\right)\left(1-\frac{s}{u}\right)\right]-\left\{ p_{2}\leftrightarrow p_{3},\text{ }\xi\rightarrow-\xi\right\} \,,\nonumber 
\end{align}
we then must set $\xi=1$.

\begin{align}\label{F_emuemu2}
&\mathcal{F}_{B;e\mu,2}^{(1)}=-\frac{1}{\epsilon^2 (1-2\epsilon)su}\left[-d\log(1432,3)\left(\frac{t(\epsilon^2-3\epsilon+2)}{u} + (4-6\epsilon +\epsilon^2)\right)\right.
&&\\
\notag\\
&\left.+d\log(1432,1)\left(\frac{ \epsilon^2 t^2-4 \epsilon s^2-\epsilon s t+4 s^2+2 s t}{2tu}\right)
-d\log(1432,2)\left(\frac{ \epsilon^2 s t+2 \epsilon^2 t^2-2 \epsilon s^2+\epsilon s t+4 s^2+2 s t}{ t u}\right)\right]+ \left\{p_2\leftrightarrow p_3\right\}\,.
\notag
\end{align}
\end{widetext}
Let us discuss the integrals that are part of the families 1432
and 1423. 
The integrals in these families can be found by considering the following exchange of external momenta,
\begin{align}
\notag
d\log(1432,i) &= d\log(1234,i)\Big|_{p_2\leftrightarrow p_4}\,,\\
d\log(1423,i) &= d\log(1243,i)\Big|_{p_2\leftrightarrow p_4}
\,,
\label{exchange1432}
\end{align}
for $i=1,2,3$.
The denominators for this family of integrals can be respectively found by performing an exchange of $p_2$ and $p_4$
and $p_2\rightarrow p_4$, $p_3\rightarrow p_2$ and $p_4 \rightarrow p_3$ in the denominators presented in Table~\ref{OneLoopProps}.

\subsubsection*{Two-loop bare form factors}
Similarly to the one-loop case, to find the two-loop form factors for the $e^+\mu^-\rightarrow e^+\mu^-$ scattering, we can take the form factors from $e^+e^-\rightarrow \mu^+\mu^-$ then perform an exchange of $s$ and $t$ and exchange $p_2$ and $p_4$ in the integral families. However, at two-loop level this is more complicated. This is due to more families being present in the two-loop form factors of the $e^+ e^-\rightarrow \mu^+ \mu^-$ scattering.
Therefore, we need to make sure we perform exchanges for all of the families.

For the planar families we have
\begin{align}
&d\log(\text{PL1432},i)= d\log(\text{PL1234},i)\Big|_{p_2\leftrightarrow p_4}
\,,&&
\label{exchanges_PL_2L}
\end{align}
for $i=1,2,\hdots,8$.

Similarly, for the non-planar families we have:
\begin{align}
 &d\log(\text{NPL1432},i)= d\log(\text{NPL1234},i)\Big|_{p_2\leftrightarrow p_4}
\,,&&
\label{exchanges_PL_2L}
\end{align}
for $i=1,2,\hdots,12$.

Similar exchanges must be performed for all families that appear in the form factors of the $e^+ e^- \rightarrow \mu^+ \mu^-$ scattering. The families that appear are listed in Appendix \ref{Canonical Families}.

The explicit expressions for the two-loop bare form factors can be found in the ancillary files.

\subsection{$e^+e^-\rightarrow e^+e^-$}
Similarly to the previous process, the $e^+e^- \rightarrow e^+e^-$ process can be derived from the helicity amplitudes of the $e^+e^-\rightarrow\mu^+\mu^-$ and $e^+\mu^-\rightarrow e^+\mu^-$.

\begin{figure}[t]
\centering
\begin{subfigure}[b]{0.2\textwidth}
        \centering
        \includegraphics[scale=0.22]{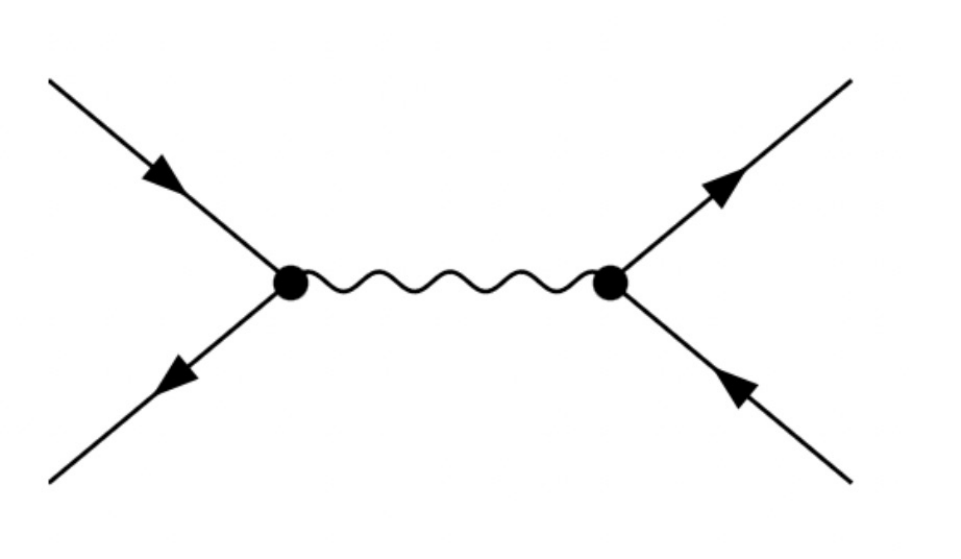}
    \end{subfigure}
    \begin{subfigure}[b]{0.2\textwidth}
        \centering
        \includegraphics[scale=0.2]{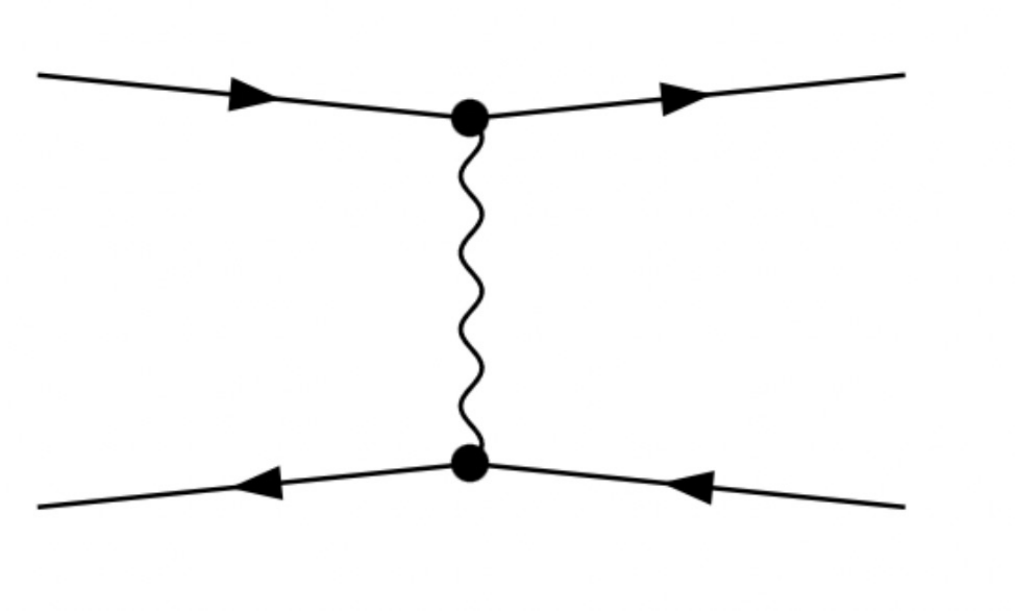}
    \end{subfigure}
\caption{The two tree-level diagrams that contribute to the $e^+e^-\rightarrow e^+e^-$ scattering. These diagrams were generated using {\sc FeynArts}~\cite{Hahn:2000kx}.
}\label{Bhabha}
\end{figure}

As all fermions are of the same flavour, additional fermionic currents are possible. Meaning that both the diagrams from the $e^+e^- \rightarrow e^+e^-$ scattering and those from the $e^+e^-\rightarrow\mu^+\mu^-$ and $e^+\mu^-\rightarrow e^+\mu^-$ contribute to the scattering amplitude. This can be seen in Fig.~\ref{Bhabha}.

We obtained the form factors for the $e^+\mu^-\rightarrow e^+\mu^-$ scattering via an exchange of the particles with momenta $p_2$ and $p_4$. However, for $e^+e^-\rightarrow e^+e^-$ if we consider the same exchange, identical fermions will be exchanged. Therefore, we must include a change of sign for all of the diagrams that can be obtained by this exchange. We are then able to combine the form factors that come from all diagrams to obtain the helicity amplitudes.

As with the $e^+e^-\rightarrow \mu^+\mu^-$ scattering, there are only 4 non-vanishing helicities. Therefore, we can present the tensor structures in the form of Eq.~(\ref{eemumuTS}). To obtain the helicity amplitudes for the $e^+e^-\rightarrow e^+e^-$ scattering, we use the following combinations of the form factors.
\begin{align}
 &\mathcal{H}^{(l)}_{ee,(-++-)}= -2 \mathcal{F}^{(l)}_{\mu\mu,1} +\left(1-x\right)\mathcal{F}^{(l)}_{\mu\mu,2}\,,&&\\&
\mathcal{H}^{(l)}_{ee,(-+-+)}= -2 \mathcal{F}^{(l)}_{\mu\mu,1} -x\mathcal{F}^{(l)}_{\mu\mu,2}-2\mathcal{F}^{(l)}_{e\mu,1}+\mathcal{F}^{(l)}_{e\mu,2}\,. \notag
\end{align}
Note that in the $(-++-)$ configuration the crossed diagrams don't contribute. This is as helicity conservation is not possible for the $e^+\mu^- \rightarrow e^+\mu^-$ scattering in this helicity configuration.

From here we can obtain the following helicity amplitudes:
\begin{align}
&\mathcal{A}^{(l)}_{ee,(\lambda_1\lambda_2\lambda_3\lambda_4)}= \mathcal{H}^{(l)}_{ee,(\lambda_1\lambda_2\lambda_3\lambda_4)}\Phi_{\mu\mu,(\lambda_1\lambda_2\lambda_3\lambda_4)}\,.&&
\end{align}
\subsection{$e^+e^-\rightarrow \gamma\gamma$}
For the $e^+e^-\rightarrow \gamma\gamma$ scattering there are 4 independent Lorentz structures that can appear in the amplitudes of contributing Feynman diagrams. 
\begin{align}\label{Compton TS}
\notag &T_{\gamma\gamma,1}=\overline{u}(p_2)\fsl{\varepsilon}_3u(p_1)\left(p_2\cdot\varepsilon_4\right)\,,&&\\
\notag&
T_{\gamma\gamma,2}=\overline{u}(p_2)\fsl{\varepsilon}_4u(p_1)\left(p_1\cdot\varepsilon_3\right)\,,&&\\
\notag&
T_{\gamma\gamma,3}=\overline{u}(p_2)\fsl{p}_3u(p_1)\left(p_1\cdot\varepsilon_3\right)\left(p_2\cdot\varepsilon_4\right)\,,&&\\
&
T_{\gamma\gamma,4}=\overline{u}(p_2)\fsl{p}_3u(p_1)\left(\varepsilon_3\cdot\varepsilon_4\right)\,,&&
\end{align}
where $\varepsilon_i$ represent the polarisation vectors of the photons with momenta $p_3$ and $p_4$. These polarisation vectors can be written in the spinor-helicity formalism as, 
\begin{align}
   \notag\varepsilon_{3,-}^\mu = \frac{\left\langle 3 | \gamma^\mu|2\right]}{\sqrt{2}\left[32\right]}\,,&&\varepsilon_{3,+}^\mu = \frac{\left\langle 2 | \gamma^\mu|3\right]}{\sqrt{2}\left\langle23\right\rangle}
   \,,&&\\
    \varepsilon_{4,-}^\mu = \frac{\left\langle 4 | \gamma^\mu|1\right]}{\sqrt{2}\left[41\right]}\,,&&\varepsilon_{4,+}^\mu = \frac{\left\langle 1 | \gamma^\mu|4\right]}{\sqrt{2}\left\langle14\right\rangle}\,. &&
\end{align}
Here we have set $p_2$ as a reference momentum for $\varepsilon_3$ and $p_1$ as a reference momentum for $\varepsilon_4$, such that $\varepsilon_3 \cdot p_2 = \varepsilon_4 \cdot p_1 =0$.

The tensor structures in Eq.~(\ref{Compton TS}) define the 4 form factors that we compute in our calculation of helicity amplitudes.

For these tensor structures there are 8 non-vanishing helicity configurations that can be seen below, 
\begin{align}
    &\notag\{-+--,-+++,-+-+,-++-,
   \\
   &\hspace{1cm} +-++,+---,+-+-,+--+\} \,.
\end{align}
In a similar way to the $e^+e^-\rightarrow\mu^+\mu^-$ process, we will present the first four helicities in our results. The other helicity configurations can be obtained by 
applying parity transformations.

The spinor helicity representation that we will use for the first four helicities respectively are:
\begin{align}
&\notag\left\{\Phi_{\gamma\gamma,(-+--)},\Phi_{\gamma\gamma,(-+++)},\Phi_{\gamma\gamma,(-+-+)},\Phi_{\gamma\gamma,(-++-)}\right\}&&\\\notag&
&&\\&\hspace{-0.5cm}=\left\{\frac{\left\langle 14 \right\rangle\left[ 24\right]\left\langle 34 \right\rangle}{\left[ 34\right]} \text{,} \frac{\left\langle 14 \right\rangle\left[ 24\right]\left[ 34 \right]}{\left\langle 34\right\rangle}\text{,}\right.
\left.\frac{\left\langle 13 \right\rangle\left[ 14\right]\left[ 24 \right]}{\left[ 13\right]}\text{,} \frac{\left\langle 14 \right\rangle\left\langle 24\right\rangle\left[ 23 \right]}{\left\langle 23\right\rangle}\right\}\,.&&
\end{align}

As with the $e^+e^-\rightarrow \mu^+\mu^-$ scattering we also have to consider pre-factors to each of the form factors when constructing the helicity amplitudes.
The sums we require are:
\begin{align}
\notag &\mathcal{H}^{(l)}_{\gamma\gamma,(-+--)}= -\frac{1}{x}\mathcal{F}^{(l)}_{\gamma\gamma,2} -\frac{(1-x)}{2x}\mathcal{F}^{(l)}_{\gamma\gamma,3}-\frac{1}{x}\mathcal{F}^{(l)}_{\gamma\gamma,4}\,,  &&\\&
\notag \mathcal{H}^{(l)}_{\gamma\gamma,(-+++)}= \frac{1}{x}\mathcal{F}^{(l)}_{\gamma\gamma,1} -\frac{(1-x)}{2x}\mathcal{F}^{(l)}_{\gamma\gamma,3}-\frac{1}{x}\mathcal{F}^{(l)}_{\gamma\gamma,4}\,,  &&\\&
\notag\mathcal{H}^{(l)}_{\gamma\gamma,(-+-+)}= \frac{(1-x)}{2x}\mathcal{F}^{(l)}_{\gamma\gamma,3}+\frac{(1-x)}{x}\mathcal{F}^{(l)}_{\gamma\gamma,4}\,,  &&\\&
\notag\mathcal{H}^{(l)}_{\gamma\gamma,(-++-)}= -\frac{1}{x}\mathcal{F}^{(l)}_{\gamma\gamma,1}+\frac{1}{x}\mathcal{F}^{(l)}_{\gamma\gamma,2}&&\\&
\hspace{2.5cm}+\frac{(1-x)}{2x}\mathcal{F}^{(l)}_{\gamma\gamma,3}+\frac{(1-x)}{x}\mathcal{F}^{(l)}_{\gamma\gamma,4}\,.
\end{align}
We then multiply these expressions by their corresponding spinor structures,
\begin{align}
&\mathcal{A}^{(l)}_{\gamma\gamma,(\lambda_1\lambda_2\lambda_3\lambda_4)}= \mathcal{H}^{(l)}_{\gamma\gamma,(\lambda_1\lambda_2\lambda_3\lambda_4)}\Phi_{\gamma\gamma,(\lambda_1\lambda_2\lambda_3\lambda_4)}\,.&&
\end{align}

From the relations discussed in this section, we obtain helicity amplitudes for all processes at tree level, one loop, and two loops. In the following section, we proceed to verify these results.

\section{Checks}\label{checks}

For all the processes we have considered we have obtained expressions that  
contain only the finite remainder and higher orders of $\epsilon$. This is already a very good indicator that our results are accurate. However, we must also confirm that higher orders in the dimensional regulator in our expressions are also valid.

To ensure that our results are consistent at higher orders in the dimensional regulator we performed checks against existing results up to the same $\epsilon$ order in corresponding QCD processes.
Namely, we can compare:
\begin{itemize}
\item $e^+e^-\rightarrow\mu^+\mu^-$ with $q\overline{q}\rightarrow Q\overline{Q}$~\cite{Caola:2021rqz,Ahmed:2019qtg},

\item $e^+\mu^-\rightarrow e^+\mu^-$ with $q\overline{Q}\rightarrow q\overline{Q}$~\cite{Caola:2021rqz},

\item $e^+e^-\rightarrow e^+e^-$ with $q\overline{q}\rightarrow q\overline{q}$~\cite{Caola:2021rqz},

\item $e^+e^-\rightarrow\gamma\gamma$ with $q\overline{q}\rightarrow gg$~\cite{Caola:2022dfa,Ahmed:2019qtg}.
\end{itemize}

Whilst this is an incredibly beneficial check, we cannot make an immediate direct comparison because the QCD processes involve additional contributing diagrams which are non-Abelian. There is an extra level of intricacy as we also have to consider the colour structure of quarks and gluons in QCD. Therefore, we need to identify the contributions to the QCD scattering amplitudes that correspond solely to the Abelian diagrams present in QED.

We will discuss each process separately as they require slightly different treatments.

\subsection{Notation}
Throughout this section, we set our notation to present the checks on our results in a concise way.
Firstly, we will represent the helicity amplitudes in form of Eqs.~\eqref{Form4fermion} and~\eqref{FormCompton}. $A_p^{(l)}\,, B_p^{(l)}\,,C_p^{(l)}$ represent combinations of Abelian diagrams at $l$ loops and $p$ denotes the process we are considering.
Additionally, to extract the relevant pieces of the QCD amplitude,\footnote{A similar organisation of Feynman diagrams present in the scattering of four fermions in QED and in QCD has been carried out in~\cite{Mandal:2022vju}.} we employ the notation,
\begin{equation} \langle\mathcal{M}_{\text{X}}^{(l)}|\mathcal{C}_j\rangle_{[k,m]}\,.
\end{equation}
The indices in this notation can be understood as:\begin{itemize}
    \item $X$ is the QCD process, 
    \item $l$ is the loop order,
    \item $j$ informs us which colour structure is being considered,
    \item $k$ tells us which power of $N_c$ we are considering,
    with $N_c$ representing the number of colours.
    \item $m$ tells us which power of $N_f$ we are considering. 
    \end{itemize}

It is also worth noting that, in order to make our checks, we must make an exchange of $x\rightarrow1-x$ as in our calculation the external particles $p_3$ and $p_4$ are swapped compared to \cite{Caola:2021rqz,Caola:2022dfa,Ahmed:2019qtg} against which we will be completing our checks. 

Additionally, in order to establish consistency between the results, we must ensure that the spinor helicity structures that appear are equivalent. Therefore, we added a helicity dependent kinematic pre-factor to each of the helicity amplitudes provided in \cite{Caola:2021rqz,Caola:2022dfa,Ahmed:2019qtg}.

Since~\cite{Ahmed:2019qtg} provides the bare helicity amplitudes, our comparison must be made with the bare helicity amplitudes derived from our bare form factors, which include poles in $\epsilon$. This contrasts with the final helicity amplitudes obtained from our UV-renormalised and IR-subtracted form factors, which contain only the finite remainder and higher orders of $\epsilon$. Due to a subtlety in the dimensional regularisation, a comparison order-by-order in $\epsilon$ between $e^+ e^- \rightarrow \mu^+ \mu^-$ and $q\overline{q}\rightarrow Q\overline{Q}$ in~\cite{Ahmed:2019qtg} is not straightforward. 
This is as the tensor structures used in the decomposition of the amplitudes have explicit dependence on the dimensional regulator $\epsilon$. This means that whilst the finite remainder\footnote{Terms proportional to $\epsilon^0$, but not higher orders of $\epsilon$.} after UV renormalisation and IR subtraction would agree, bare helicity amplitudes do not. 
The same issue was reported in~\cite{Caola:2021rqz} as the authors of this reference performed similar cross-checks.

The bare helicity amplitudes for $q\overline{q} \rightarrow gg$ are also provided in \cite{Ahmed:2019qtg}, therefore we are able to compare our bare helicity amplitudes of $e^+e^-\rightarrow \gamma\gamma$ from our bare form factors with the helicity amplitudes presented in \cite{Ahmed:2019qtg}. The same subtlety in the dimensional regularisation does not appear to affect the results of this process. In the decomposition of $q\overline{q} \rightarrow gg$, Ref.~\cite{Ahmed:2019qtg} uses 5 alternative tensors structures as opposed to the 4 tensor structures that we use. However, we were able to find full agreement for the bare helicity amplitudes at tree-level, one- and two-loop at all orders in $\epsilon$ that are presented.

\subsection{Checks on $e^+e^-\rightarrow\mu^+\mu^-$}
\label{checks_eemumu}

To perform our checks for the $e^+e^-\rightarrow\mu^+\mu^-$ scattering amplitude we make a comparison against the results obtained in \cite{Caola:2021rqz}. 
The colour structures we have to consider for this process are:
\begin{align}
\notag &\mathcal{C}_{1}=\delta_{i_1i_3}\delta_{i_2i_4}\,,&&\\
 &\mathcal{C}_{2}=\delta_{i_1i_2}\delta_{i_3i_4} \,,\label{Colourseemumu}&&
\end{align}
 where $\delta_{ij}$ is the colour delta function.

The helicity amplitudes for this process take the form of Eq.~\eqref{Form4fermion}. To verify our results, we make comparisons between QED and QCD using the following relations:
\begin{itemize}
    \item Tree-Level
    \begin{equation} 
    A_{\mu\mu}^{(0)}=-2\times\langle\mathcal{M}_{Q\overline{Q}}^{(0)}|\mathcal{C}_2\rangle_{[-1,0]}\,.
\end{equation}
\item One-Loop
\begin{subequations}
    \begin{equation} A_{\mu\mu}^{(1)}=2\times\langle\mathcal{M}_{Q\overline{Q}}^{(1)}|\mathcal{C}_2\rangle_{[-2,0]}\,,
\end{equation}
  \begin{equation} B_{\mu\mu}^{(1)}=-2\times\langle\mathcal{M}_{Q\overline{Q}}^{(1)}|\mathcal{C}_2\rangle_{[-1,1]}\,.
\end{equation}
\end{subequations}
\item Two-Loop
\begin{subequations}
    \begin{equation} A_{\mu\mu}^{(2)}=-2\times\langle\mathcal{M}_{Q\overline{Q}}^{(2)}|\mathcal{C}_2\rangle_{[-3,0]}\,,
\end{equation}
  \begin{equation}  B_{\mu\mu}^{(2)}=2\times\langle\mathcal{M}_{Q\overline{Q}}^{(2)}|\mathcal{C}_2\rangle_{[-2,1]}\,,
\end{equation}
 \begin{equation}  C_{\mu\mu}^{(2)}=-2\times\langle\mathcal{M}_{Q\overline{Q}}^{(2)}|\mathcal{C}_2\rangle_{[-1,2]}\,.
\end{equation}
\end{subequations}
\end{itemize}
Using these comparisons, we find complete agreement with the QCD results in \cite{Caola:2021rqz} at all orders in $\epsilon$ at  one and two loops. Whilst \cite{Ahmed:2019qtg} provides expressions for the bare helicity amplitudes relevant for this process, we are not able to make a direct comparison due to differences in the dimensional regularisation schemes that are used. This discrepancy is also noted in \cite{Caola:2021rqz}. In Sec.~\ref{IterativeIR} we aim to clarify the differences that are observed.

\subsection{Checks on $e^+\mu^-\rightarrow e^+\mu^-$}

To perform our checks for the $e^+\mu^-\rightarrow e^+\mu^-$ scattering amplitude, we make a comparison against the results obtained in \cite{Caola:2021rqz}. 
Here, the colour structures we have to consider for this process are:
\begin{align}
\notag &\mathcal{C}_{1}=\delta_{i_1i_4}\delta_{i_2i_3}\,,&&\\
 &\mathcal{C}_{2}=\delta_{i_1i_2}\delta_{i_3i_4} \,.\label{Colourseemumu}&&
\end{align}
Similar to $e^+e^-\rightarrow\mu^+\mu^-$, we make the following comparisons:
\begin{itemize}
    \item Tree-Level
    \begin{equation} A_{e\mu}^{(0)}=-2\times\langle\mathcal{M}_{q\overline{Q}}^{(0)}|\mathcal{C}_1\rangle_{[-1,0]}\,.
\end{equation}
\item One-Loop
\begin{subequations}
    \begin{equation} A_{e\mu}^{(1)}=2\times\langle\mathcal{M}_{q\overline{Q}}^{(1)}|\mathcal{C}_1\rangle_{[-2,0]}\,,
\end{equation}
  \begin{equation} B_{e\mu}^{(1)}=-2\times\langle\mathcal{M}_{q\overline{Q}}^{(1)}|\mathcal{C}_1\rangle_{[-1,1]}\,.
\end{equation}
\end{subequations}
\item Two-Loop
\begin{subequations}
    \begin{equation} A_{e\mu}^{(2)}=-2\times\langle\mathcal{M}_{q\overline{Q}}^{(2)}|\mathcal{C}_1\rangle_{[-3,0]}\,,
\end{equation}
  \begin{equation} B_{e\mu}^{(2)}=2\times\langle\mathcal{M}_{q\overline{Q}}^{(2)}|\mathcal{C}_1\rangle_{[-2,1]}\,,
\end{equation}
 \begin{equation}  C_{e\mu}^{(2)}=-2\times\langle\mathcal{M}_{q\overline{Q}}^{(2)}|\mathcal{C}_1\rangle_{[-1,2]}\,.
\end{equation}
\end{subequations}
\end{itemize}
Using these comparisons, we find complete agreement with the QCD results in \cite{Caola:2021rqz} at one and two loops.

\subsection{Checks on $e^+e^-\rightarrow e^+e^-$}
To perform our checks for the $e^+e^-\rightarrow e^+e^-$ scattering amplitude, we make a comparison against the results obtained for the same flavour process in \cite{Caola:2021rqz}. 
The colour structures we have to consider for this process are also the same as those considered in Eq. (\ref{Colourseemumu}).
The helicity amplitudes for this process also take the form of Eq. \eqref{Form4fermion}. However, our checks are slightly different as both colour structures need to be considered, as seen below:
\begin{itemize}
    \item Tree-Level
    \begin{equation} 
A_{ee}^{(0)}=-2\times\left(\langle\mathcal{M}_{q\overline{q}}^{(0)}|\mathcal{C}_1\rangle_{[-1,0]}+\langle\mathcal{M}_{q\overline{q}}^{(0)}|\mathcal{C}_2\rangle_{[-1,0]}\right)\,.
\end{equation}
\item One-Loop
\begin{subequations}
    \begin{equation} A_{ee}^{(1)}=2\times\left(\langle\mathcal{M}_{q\overline{q}}^{(1)}|\mathcal{C}_1\rangle_{[-2,0]}+\langle\mathcal{M}_{q\overline{q}}^{(1)}|\mathcal{C}_2\rangle_{[-2,0]}\right)\,,
\end{equation}
  \begin{equation} B_{ee}^{(1)}=-2\times\left(\langle\mathcal{M}_{q\overline{q}}^{(1)}|\mathcal{C}_1\rangle_{[-1,1]}+\langle\mathcal{M}_{q\overline{q}}^{(1)}|\mathcal{C}_2\rangle_{[-1,1]}\right)\,.
\end{equation}
\end{subequations}
\item Two-Loop
\begin{subequations}
    \begin{equation} A_{ee}^{(2)}=-2\times\left(\langle\mathcal{M}_{q\overline{q}}^{(2)}|\mathcal{C}_1\rangle_{[-3,0]}+\langle\mathcal{M}_{q\overline{q}}^{(2)}|\mathcal{C}_2\rangle_{[-3,0]}\right)\,,
\end{equation}
  \begin{equation} B_{ee}^{(2)}=2\times\left(\langle\mathcal{M}_{q\overline{q}}^{(2)}|\mathcal{C}_1\rangle_{[-2,1]}+\langle\mathcal{M}_{q\overline{q}}^{(2)}|\mathcal{C}_2\rangle_{[-2,1]}\right)\,,
\end{equation}
 \begin{equation}  C_{ee}^{(2)}=-2\times\left(\langle\mathcal{M}_{q\overline{q}}^{(2)}|\mathcal{C}_1\rangle_{[-1,2]}+\langle\mathcal{M}_{q\overline{q}}^{(2)}|\mathcal{C}_2\rangle_{[-1,2]}\right)\,.
\end{equation}
\end{subequations}
\end{itemize}
Using these comparisons, we find complete agreement with the QCD results in QCD results in \cite{Caola:2021rqz} at all orders in $\epsilon$ at one and two loops.

\subsection{Checks on $e^+e^-\rightarrow\gamma\gamma$}

To perform our checks for the $e^+e^-\rightarrow\gamma\gamma$ scattering amplitude we make a comparison against the results obtained in~\cite{Ahmed:2019qtg,Caola:2022dfa}. 
The colour structures we have to consider for this process are:
\begin{align}
\notag &\mathcal{C}_{1}=\left(T^{a_3}T^{a_4}\right)_{i_2i_1}\,,&&\\
 \notag &\mathcal{C}_{2}=\left(T^{a_4}T^{a_3}\right)_{i_2i_1}\,,&&\\
 &\mathcal{C}_{3}=\delta^{a_3 a_4}\delta_{i_2 i_1}\,.&&
\end{align}
Complications arise when making comparisons in this process, particularly when comparing the terms proportional to $N_f$. This is as we need to consider linear combinations of pieces of the QCD amplitude that correspond to different colour structures and, in some cases, orders of $N_c$ also. We have to consider these combinations to ensure that all of the Abelian diagrams are included and that each diagram only contributes once.
 
In order to verify our results
for the helicity amplitudes of this process, we cross-checked against QCD results in the following way:
\begin{itemize}
    \item Tree-Level
    \begin{equation} A_{\gamma\gamma}^{(0)}=2\times\left(\langle\mathcal{M}_{gg}^{(0)}|\mathcal{C}_1\rangle_{[0,0]}+\langle\mathcal{M}_{gg}^{(0)}|\mathcal{C}_2\rangle_{[0,0]}\right)\,.
\end{equation}
\item One-Loop
    \begin{equation} A_{\gamma\gamma}^{(1)}=-2\times\left(\langle\mathcal{M}_{gg}^{(1)}|\mathcal{C}_1\rangle_{[-1,0]}+\langle\mathcal{M}_{gg}^{(1)}|\mathcal{C}_2\rangle_{[-1,0]}\right)\,.
\end{equation}
\item Two-Loop
\begin{subequations}
    \begin{equation} A_{\gamma\gamma}^{(2)}=2\times\left(\langle\mathcal{M}_{gg}^{(2)}|\mathcal{C}_1\rangle_{[-2,0]}+\langle\mathcal{M}_{gg}^{(2)}|\mathcal{C}_2\rangle_{[-2,0]}\right)\,,
\end{equation}
  \begin{align} 
\notag &B_{\gamma\gamma}^{(2)}=-2\times\left(\langle\mathcal{M}_{gg}^{(2)}|\mathcal{C}_1\rangle_{[-1,1]}+\langle\mathcal{M}_{gg}^{(2)}|\mathcal{C}_2\rangle_{[-1,1]}\right.&&\\
&\hspace{3cm}\left.+2\times\langle\mathcal{M}_{gg}^{(2)}|\mathcal{C}_3\rangle_{[-2,1]}\right)\,.&&
\end{align}
\end{subequations}
\end{itemize}
Note that for the case of $B_{\gamma\gamma}^{(2)}$, we have to use a linear combination of the colours with differing coefficients. We use this linear combination to ensure that we only extract one contribution from each Abelian diagram from the QCD amplitude.

Using these comparisons, we find complete agreement with the QCD results in~\cite{Caola:2022dfa} at all orders in $\epsilon$ at one and two loops.
Additionally, using the same comparisons on bare helicity amplitudes provided in~\cite{Ahmed:2019qtg},
we were able to find agreement at all orders in $\epsilon$ at one and two loops as well.

\subsection{Iterative IR structure of the amplitudes}\label{IterativeIR}

In Sec.~\ref{checks_eemumu}, we observed that a direct comparison between the bare helicity amplitudes reported in~\cite{Ahmed:2019qtg} and those computed in the present manuscript is only meaningful at the level of the finite remainder.
In this section, we illustrate the iterative IR structure of the amplitudes, particularly when the logarithm of the amplitudes is considered. 
This approach, that also leads to the finite remainder, resolves discrepancies at the $\epsilon$ poles arising from differences in regularisation schemes.


From relations~\eqref{eq:FF_mumu}, 
we express our helicity amplitudes in terms of master integrals admitting a $d\log$ representation. 
By inspecting the expression of these amplitudes, 
we identify a drop in the maximal transcendental degree in terms proportional to $N_f$ and $N_f^2$.
In detail, we observe that, 
\begin{align}
\mathcal{H}_{\mu\mu\left(\lambda_{1}\lambda_{2}\lambda_{3}\lambda_{4}\right)}^{\left(2\right)}&\sim\tilde{A}_{\mu\mu}^{\left(2\right)}+\epsilon\,N_{f}\,\tilde{B}_{\mu\mu}^{\left(2\right)}+\epsilon^{2}\,N_{f}^{2}\,\tilde{C}_{\mu\mu}^{\left(2\right)}\,.
\end{align}
After performing UV renormalisation on our expressions, following~\eqref{eq:uv_renormalisation},
and comparing with the results in~\cite{Ahmed:2019qtg} once their expressions are renormalised, 
we find complete agreement in terms proportional to $N_f^2$.
For terms proportional to $N_f$ and $N_f^0$, 
we observe agreement up to the $\epsilon^{-1}$ and $\epsilon^{-2}$ poles, respectively. 
At higher orders in $\epsilon$, both results exhibit the same functional structure when considering the maximal transcendental weight. 
This indicates to differences in regularisation schemes, as mentioned above.

To address this discrepancy, we can examine the iterative IR structure of the amplitude by taking the logarithm of our renormalised amplitudes,
\begin{align}
\log&\,\mathcal{M}_{\lambda_{1}\lambda_{2}\lambda_{3}\lambda_{4}}=\left(\frac{\alpha}{2\pi}\right)\bigg[\mathcal{M}_{\lambda_{1}\lambda_{2}\lambda_{3}\lambda_{4}}^{\left(1\right)}\\&+\left(\frac{\alpha}{2\pi}\right)\left(\mathcal{M}_{\lambda_{1}\lambda_{2}\lambda_{3}\lambda_{4}}^{\left(2\right)}-\frac{1}{2}\left(\mathcal{M}_{\lambda_{1}\lambda_{2}\lambda_{3}\lambda_{4}}^{\left(1\right)}\right)^{2}\right)\bigg]+\mathcal{O}\left(\alpha^{3}\right)\,,
\notag
\end{align}
with, 
\begin{align}
\mathcal{M}_{\lambda_{1}\lambda_{2}\lambda_{3}\lambda_{4}}^{\left(l\right)}&=\frac{\mathcal{A}_{\lambda_{1}\lambda_{2}\lambda_{3}\lambda_{4}}^{\left(l\right)}}{\mathcal{A}_{\lambda_{1}\lambda_{2}\lambda_{3}\lambda_{4}}^{\left(0\right)}}\,.
\end{align}
By focusing on the $\alpha^2$ contribution, 
we obtain for the four-fermion scattering in both QED and QCD (considering only the sum of Abelian diagrams),
\begin{align}
\label{eq:log_mumu}
\mathcal{M}_{\mu\mu\left(\lambda_{1}\lambda_{2}\lambda_{3}\lambda_{4}\right)}^{\left(2\right)} & -\frac{1}{2}\left(\mathcal{M}_{\mu\mu\left(\lambda_{1}\lambda_{2}\lambda_{3}\lambda_{4}\right)}^{\left(1\right)}\right)^{2}\\
= & -\frac{1}{\epsilon^{3}}N_{f}\nonumber \\
 & +\frac{2}{3\epsilon^{2}}N_{f}\left[-\frac{2}{3}+\left(L_{34}+L_{23}-L_{24}\right)\right]\nonumber \\
 & +\frac{1}{\epsilon}\left[\gamma_{1}^{q}+\frac{\gamma_{1}^{\text{cusp}}}{2}\left(L_{34}+L_{23}-L_{24}\right)\right]\nonumber \\
 & +E_{\mu\mu\left(\lambda_{1}\lambda_{2}\lambda_{3}\lambda_{4}\right)}+\mathcal{O}\left(\epsilon\right)\,.
 \nonumber 
\end{align}
This result confirms that these scheme-dependent differences are resolved for four-fermion scattering processes in the chosen framework.

The very similar behaviour is observed for the scattering processes $q\overline{q}\rightarrow gg$
and $e^+e^-\rightarrow \gamma\gamma$,
\begin{align}
\label{eq:log_gg}
\mathcal{M}_{\gamma\gamma\left(\lambda_{1}\lambda_{2}\lambda_{3}\lambda_{4}\right)}^{\left(2\right)} & -\frac{1}{2}\left(\mathcal{M}_{\gamma\gamma\left(\lambda_{1}\lambda_{2}\lambda_{3}\lambda_{4}\right)}^{\left(1\right)}\right)^{2}\\
= & -\frac{1}{2\epsilon^{3}}N_{f}\nonumber \\
 & +\frac{1}{3\epsilon^{2}}N_{f}\left[\frac{2}{3}N_{f}+\left(-\frac{2}{3}+L_{34}\right)\right]\nonumber \\
 & +\frac{1}{2\epsilon}\left[\gamma_{1}^{g}+\gamma_{1}^{q}+\frac{\gamma_{1}^{\text{cusp}}}{2}L_{34}\right]\nonumber \\
 & +E_{\gamma\gamma\left(\lambda_{1}\lambda_{2}\lambda_{3}\lambda_{4}\right)}+\mathcal{O}\left(\epsilon\right)\,,\nonumber 
 \nonumber 
\end{align}
In Eqs.~\eqref{eq:log_mumu} and~\eqref{eq:log_gg}, 
$E$ contains information on the finite remainder, which depends on the helicity configuration.

The observation that we find the same iterative IR structure in both our results and the Abelian contributions from the results presented in~\cite{Ahmed:2019qtg}, is confirmation that the discrepancies in our results stem only from the difference in regularisation schemes.

\section{Discussion and outlook}
\label{conclusion}

In this paper, we analytically calculated one- and two-loop helicity amplitudes for the massless QED scattering processes:
\begin{align}
\notag& e^+e^-  \rightarrow \mu^+\mu^-\,,&&\\
\notag &e^+\mu^-  \rightarrow e^+\mu^-\,,&&\\
\notag &e^+e^-  \rightarrow e^+e^-\,,&&\\
\notag& e^+e^- \rightarrow \gamma\gamma\,,&&
\end{align}
at higher orders in the dimensional regulator $\epsilon$, by decomposing these physical amplitudes in terms of four-dimensional form factors. We performed this decomposition by profiting from the dimensionality of the external momenta and closely adopted the ’t Hooft-Veltman regularisation scheme. Our results for non-vanishing helicity amplitudes,
after performing UV renormalisation and IR subtraction, are expressed in terms of generalised polylogarithms up to transcendental weight six.

In addition to the standard approach to calculate multi-loop scattering amplitudes, we proposed an algorithm at integrand level, based on matrix transformations, to organise multi-loop Feynman diagrams into families. We motivated this method to simplify the complexity in the solution of IBP systems, by allowing us to identify a minimal set of master integrals that we chose them to admit a $d\log$ representation.

We observed that, by carefully extracting Abelian
contributions from the QCD counterpart processes,
we were able to verify our QED results.

With the explicit evaluation of these scattering amplitudes, there are several directions. 
The techniques we have employed are directly applicable to the calculation of three-loop scattering amplitudes, which will complete the necessary ingredients for theoretical predictions at N$^3$LO.
Additionally, we expect that our algorithm for grouping Feynman diagrams will be applicable for other physical processes involving internal massive particles.
Furthermore, we foresee the four-dimensional tensor decomposition to facilitate efficient calculations of physical processes of interest at low energies.

\section*{Acknowledgements}

We thank Giulio Gambuti and Lorenzo Tancredi for sharing updated results of Ref.~\cite{Caola:2021rqz} with us.
We also wish to thank Giulio Crisanti, Manoj Mandal, Pierpaolo Mastrolia, Jonathan Ronca and Sid Smith for collaboration on closely related projects, and Taushif Ahmed and Einan Gardi for useful discussions.
This work is supported by the Leverhulme Trust, LIP-2021-01.
We acknowledge the Mainz Institute for Theoretical Physics (MITP) of the Cluster of Excellence $\text{PRISMA}^+$ (Project ID 390831469), for its hospitality.

\newpage

\appendix
\begin{widetext}
\section{Results obtained via Matrix-Based Method}\label{Grouping Results}
Utilising the algorithm presented in Sec.~\ref{groups} 
for the $e^+ e^- \rightarrow \mu^+ \mu^-$ process, we find shifts for all diagrams apart from 2 (for which there was no possible shift of loop momenta). These 2 diagrams were then defined as additional parent diagrams, giving 18 families altogether. Table~\ref{fig:Families emu} shows how many diagrams, including the parent topology, appear in each family.

Analogously, for the $e^+ e^- \rightarrow \gamma\gamma$ process, 
we find that there were shifts for all diagrams into 20 families. Table~\ref{fig:Families Compton} shows matching information for this process.
\begin{table}[h]
\begin{minipage}{0.45\linewidth}
\center \begin{tabular}{|cll|}
\hline
\multicolumn{2}{|c|}{\textbf{Parent Diagram}}     & \textbf{\# Diagrams}                                                                                                                                                                                               \\ \hline\hline
\multicolumn{1}{|c|}{\multirow{12}{*}{\rotatebox[origin=c]{90}{\textbf{Planar}}}}    & \multicolumn{1}{l|}{Diagram 16} & 5                                                                                                                                      \\ \cline{2-3} 
\multicolumn{1}{|c|}{}                            & \multicolumn{1}{l|}{Diagram 19} & 2                                                                                                                                              \\ \cline{2-3} 
\multicolumn{1}{|c|}{}                            & \multicolumn{1}{l|}{Diagram 20} & 30 \\ \cline{2-3} 
\multicolumn{1}{|c|}{}                            & \multicolumn{1}{l|}{Diagram 22} & 5                                                                                                                                     \\ \cline{2-3} 
\multicolumn{1}{|c|}{}                            & \multicolumn{1}{l|}{Diagram 23} & 2                                                                                                                                                     \\ \cline{2-3} 
\multicolumn{1}{|c|}{}                            & \multicolumn{1}{l|}{Diagram 25} & 6                                                                                                                                 \\ \cline{2-3} 
\multicolumn{1}{|c|}{}                            & \multicolumn{1}{l|}{Diagram 26} & 4                                                                                                                                          \\ \cline{2-3} 
\multicolumn{1}{|c|}{}                            & \multicolumn{1}{l|}{Diagram 27} & 4                                                                                                                                        \\ \cline{2-3} 
\multicolumn{1}{|c|}{}                            & \multicolumn{1}{l|}{Diagram 28} & 2                                                                                                                                               \\ \cline{2-3} 
\multicolumn{1}{|c|}{}                            & \multicolumn{1}{l|}{Diagram 29} & 1                                                                                                                                                   \\ \cline{2-3} 
\multicolumn{1}{|c|}{}                            & \multicolumn{1}{l|}{Diagram 31} & 1                                                                                                                                                      \\ \cline{2-3} 
\multicolumn{1}{|c|}{}                            & \multicolumn{1}{l|}{Diagram 32} & 1                                                                                                                                                     \\ \hline\hline
\multicolumn{1}{|c|}{\multirow{6}{*}{\rotatebox[origin=c]{90}{\textbf{Non-Planar}}}}  & \multicolumn{1}{l|}{Diagram 17} & 1                                                                                                                                                    \\ \cline{2-3} 
\multicolumn{1}{|c|}{}                            & \multicolumn{1}{l|}{Diagram 18} & 1                                                                                                                                                      \\ \cline{2-3} 
\multicolumn{1}{|c|}{}                            & \multicolumn{1}{l|}{Diagram 30} & 1                                                                                                                                                     \\ \cline{2-3} 
\multicolumn{1}{|c|}{}                            & \multicolumn{1}{l|}{Diagram 33} & 1                                                                                                                                                   \\ \cline{2-3} 
\multicolumn{1}{|c|}{}                            & \multicolumn{1}{l|}{Diagram 42} & 1                                                                                                                                                     \\ \cline{2-3} 
\multicolumn{1}{|c|}{}                            & \multicolumn{1}{l|}{Diagram 49} & 1                                                                                                                                                   \\ \hline
\end{tabular}
\caption{Families for $e^+e^-\rightarrow \mu^+ \mu^-$.}
    \label{fig:Families emu}
    \end{minipage}
\begin{minipage}{0.45\linewidth}
    \center \begin{tabular}{|cll|}
\hline
\multicolumn{2}{|c|}{\textbf{Parent Diagram}}     & \textbf{\# Diagrams}                                                                                                                                                                                               \\ \hline\hline
\multicolumn{1}{|c|}{\multirow{12}{*}{\rotatebox[origin=c]{90}{\textbf{Planar}}}}    & \multicolumn{1}{l|}{Diagram 49} & 9                                                                                                                                      \\ \cline{2-3} 
\multicolumn{1}{|c|}{}                            & \multicolumn{1}{l|}{Diagram 52} & 6                                                                                                                                              \\ \cline{2-3} 
\multicolumn{1}{|c|}{}                            & \multicolumn{1}{l|}{Diagram 53} & 25 \\ \cline{2-3} 
\multicolumn{1}{|c|}{}                            & \multicolumn{1}{l|}{Diagram 58} & 10                                                                                                                                     \\ \cline{2-3} 
\multicolumn{1}{|c|}{}                            & \multicolumn{1}{l|}{Diagram 63} & 5                                                                                                                                                     \\ \cline{2-3} 
\multicolumn{1}{|c|}{}                            & \multicolumn{1}{l|}{Diagram 68} & 14                                                                                                                                 \\ \cline{2-3} 
\multicolumn{1}{|c|}{}                            & \multicolumn{1}{l|}{Diagram 69} & 7                                                                                                                                          \\ \cline{2-3} 
\multicolumn{1}{|c|}{}                            & \multicolumn{1}{l|}{Diagram 70} & 7                                                                                                                                        \\ \cline{2-3} 
\multicolumn{1}{|c|}{}                            & \multicolumn{1}{l|}{Diagram 71} & 2                                                                                                                                               \\ \cline{2-3} 
\multicolumn{1}{|c|}{}                            & \multicolumn{1}{l|}{Diagram 73} & 8                                                                                                                                                   \\ \cline{2-3} 
\multicolumn{1}{|c|}{}                            & \multicolumn{1}{l|}{Diagram 74} & 9                                                                                                                                                      \\ \cline{2-3} 
\multicolumn{1}{|c|}{}                            & \multicolumn{1}{l|}{Diagram 78} & 2            
                                                                     \\ \cline{2-3} 
\multicolumn{1}{|c|}{}                            & \multicolumn{1}{l|}{Diagram 80} & 9     
                                                                     \\ \cline{2-3} 
\multicolumn{1}{|c|}{}                            & \multicolumn{1}{l|}{Diagram 84} & 8     
\\ \hline\hline
\multicolumn{1}{|c|}{\multirow{6}{*}{\rotatebox[origin=c]{90}{\textbf{Non-Planar}}}}  & \multicolumn{1}{l|}{Diagram 47} & 6                                                                                                                                                    \\ \cline{2-3} 
\multicolumn{1}{|c|}{}                            & \multicolumn{1}{l|}{Diagram 50} & 4                                                                                                                                                      \\ \cline{2-3} 
\multicolumn{1}{|c|}{}                            & \multicolumn{1}{l|}{Diagram 51} & 7                                                                                                                                                     \\ \cline{2-3} 
\multicolumn{1}{|c|}{}                            & \multicolumn{1}{l|}{Diagram 59} & 4                                                                                                                                                   \\ \cline{2-3} 
\multicolumn{1}{|c|}{}                            & \multicolumn{1}{l|}{Diagram 72} & 3                                                                                                                                                     \\ \cline{2-3} 
\multicolumn{1}{|c|}{}                            & \multicolumn{1}{l|}{Diagram 79} & 3                                                                                                                                                   \\ \hline
\end{tabular}
\caption{Families for $e^+e^-\rightarrow \gamma\gamma$.}
    \label{fig:Families Compton}
    \end{minipage}
\end{table}

The Feynman diagrams for the parent topologies that define each family can be found in Figs. \ref{emu_diags} and \ref{compt_diags}.
These tables summarise information about the families that the integrands are grouped into (see Sec.~\ref{canonical} 
for the definition of the families).
\begin{figure}[t]
    \centering
    \begin{subfigure}[b]{0.45\textwidth}
        \centering
        \includegraphics[scale=0.45]{2L_Box.pdf}
        \caption{Diagrams: \textbf{20}, 23, 29, 32.}
    \end{subfigure}
    \hfill
    \begin{subfigure}[b]{0.45\textwidth}
        \centering
        \includegraphics[scale=0.45]{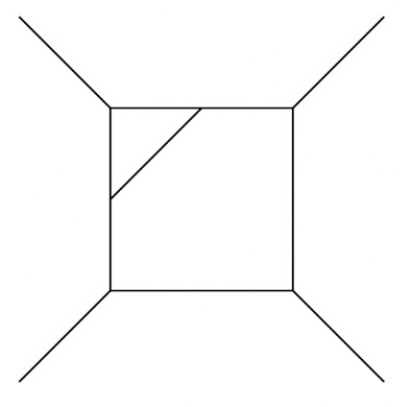}
        \caption{Diagrams: \textbf{16}, 19, 22, 25, 26, 27, 28, 31.}
    \end{subfigure}
     \hfill
    \begin{subfigure}[b]{0.45\textwidth}
        \centering
        \includegraphics[scale=0.45]{2L_NP.pdf}
        \caption{Diagrams: \textbf{17}, 18, 30, 33.}
    \end{subfigure}
         \hfill
    \begin{subfigure}[b]{0.45\textwidth}
        \centering
        \includegraphics[scale=0.45]{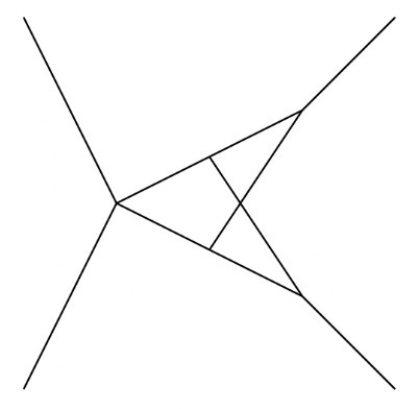}
        \caption{Diagrams: \textbf{42}, 49.}
    \end{subfigure}
    \caption{Parent topologies present in the two-loop scattering amplitude $e^+ e^- \rightarrow \mu^+\mu^-$. Here and in Fig.~\ref{compt_diags}, the numbers in bold are the diagrams represented by the Figure in the standard momenta ordering, the others can be found via exchanges of external momenta.}\label{emu_diags}
\end{figure}
\begin{figure}[t]
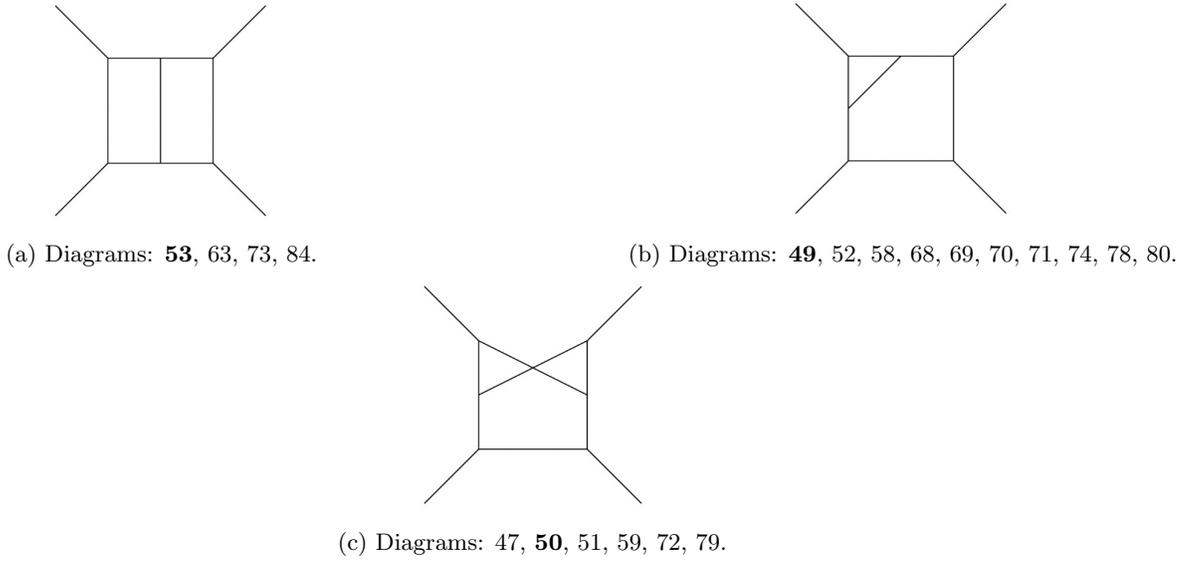

    \centering
    \begin{subfigure}[b]{0.45\textwidth}
        \centering
        \includegraphics[scale=0.45]{2L_Box.pdf}
        \caption{Diagrams: \textbf{53}, 63, 73, 84.}
    \end{subfigure}
    \hfill
    \begin{subfigure}[b]{0.45\textwidth}
        \centering
        \includegraphics[scale=0.45]{2L_Triangle.pdf}
        \caption{Diagrams: \textbf{49}, 52, 58, 68, 69, 70, 71, 74, 78, 80.}
    \end{subfigure}
     \hfill
    \begin{subfigure}[b]{0.45\textwidth}
        \centering
        \includegraphics[scale=0.45]{2L_NP.pdf}
        \caption{Diagrams: 47, \textbf{50}, 51, 59, 72, 79.}
    \end{subfigure}
    \caption{Parent topologies present in the two-loop $e^+ e^- \rightarrow \gamma\gamma$. }\label{compt_diags}
\end{figure}
 \end{widetext}
\section{Canonical integral families}\label{Canonical Families}

\subsection*{Canonical families for $e^+e^-\rightarrow\mu^+\mu^-$}

For the one-loop case, the integrals 
that appear in families 1234 and 1243 
provide a sufficient basis of MIs to express all scalar integrals that appear.

At two-loop, we are able to express all integrals in terms of 7 integral families. As mentioned in Sec.~\ref{canonical}, we prioritise planar integrals over non-planar integrals. 
This allows us to reduce from the 120 canonical integrals (from 12 families) to just 35 MIs that stem from 7 families.

Namely, we can reduce our form factors to
integrals appearing in the planar families: PL1234, PL1243, PL1342, PL1432,
and in the non-planar families: NPL1234, NPL1243, NPL1432.

\subsection*{Canonical families for $e^+e^-\rightarrow\gamma\gamma$}
As with the $e^+e^-\rightarrow\mu^+\mu^-$ process in the one-loop case, the integral families 1234 and 1243 provide a sufficient basis of MIs to express all scalar integrals that appear.

For the two-loop case of this process, we are able to express all integrals in terms of 39 MIs that originate from 9 integral families. These families are comprised of all of the 6 planar families and 3 non-planar families.

The planar families are: PL1234, PL1243, PL1324, PL1342, PL1423, PL1432, 
as well as the non-planar families:
NPL1234, NPL1243, NPL1342.

The denominators for the families required for each process can be obtained using the propagators in Table~\ref{Props} and performing any necessary exchanges of momenta.

The definitions of these integrals in terms of scalar integrals can be found within the ancillary files.

\section{Details on IR subtraction}
\label{Coeffs}

For $e^+ e^- \rightarrow \mu^+ \mu^-$ 
and  $e^+ e^- \rightarrow \gamma \gamma$,
we can, respectively, use the following formulae for the elements of the anomalous dimension matrix $\Gamma$:
\begin{align}
\Gamma_{\mu\mu;\,n} &= -2\gamma_n^{\text{cusp}}\left(L_{34}+L_{23}-L_{24}\right)+4\gamma_n^q\,,
\\
\Gamma_{\gamma\gamma;\, n} &= -\gamma_n^{\text{cusp}}\left(L_{34}\right)+2\gamma_n^q+2\gamma_n^g\,,
\end{align}
and
\begin{align}
\Gamma_{\mu\mu;\,n}' &= \frac{\partial\Gamma_{\mu\mu;\,n}}{\partial \log(\mu)}= -4\gamma_n^{\text{cusp}}\,,\\
\Gamma_{\gamma\gamma;\, n}' &= \frac{\partial\Gamma_{\gamma\gamma;\, n}}{\partial \log(\mu)}= -2\gamma_n^{\text{cusp}}\,.
\end{align}
Below, we list the anomalous dimension coefficients for the relevant particles in our processes. We obtained these values considering the coefficients of $\left(\frac{\alpha_S^{(n_l)}}{2\pi}\right)^{n+1}$ in the expansions given in \cite{Barnreuther:2013qvf,Ahrens:2012qz,Becher:2009kw,Kidonakis:2009zc}, once we abelianised these expansions by setting $C_A\to0$, $C_F\to1$, and $T_F\to1$.
\\
\\
\textbf{The cusp anomalous dimension}
\begin{flalign}\label{cuspAD}
\notag &\gamma_0^{\text{cusp}}=2\,, &&\\
&\gamma_1^{\text{cusp}}=-\frac{20}{9}N_f\,.&&
\end{flalign}
\textbf{The quark anomalous dimension}
\begin{flalign}\label{quarkAD}
\notag &\gamma_0^{q}=-\frac{3}{2}\,, &&\\
&\gamma_1^{q}=\left(\frac{65}{54}+\frac{\pi^2}{6}\right)N_f+\left(\frac{\pi^2}{2}-6\zeta_3-\frac{3}{8}\right)\,.&&
\end{flalign}
\\\textbf{The gluon anomalous dimension}
\begin{flalign}\label{gluonAD}
\notag &\gamma_0^{g}=-\beta_0\,, &&\\
&\gamma_1^{g}=-\beta_1\,.&&
\end{flalign}

\section{Organisation of the ancillary files}
\label{Files}

In this appendix we outline the organisation of the ancillary files included on the arXiv submission of this paper:

\begin{itemize}

\item \verb"Bare_FFs/": we provide the decompositions of the bare form factors for the processes $e^+e^-\rightarrow \mu^+\mu^-$, $e^+\mu^-\rightarrow e^+\mu^-$ and $e^+e^-\rightarrow\gamma\gamma$ at tree level, one-loop, and two-loop level in terms of master integrals. We also provide additional files where we express these master integrals in terms of GPLs.

\item \verb"Results/": in this folder we include the decompositions of the UV renormalised and IR subtracted form factors, which are free from poles, for each process at each loop order.

\item \verb"Helicity_Amplitudes/": we supply the non-vanishing helicity amplitudes, that are free from poles, for all processes across tree level, one-loop and two-loop orders.

\item \verb"Families/": we provide the master integrals in the form of scalar integrals~\eqref{scalar integral}, for all families. The denominators of these scalar integrals can be deduced from Tables~\ref{OneLoopProps} and~\ref{Props}.
\end{itemize}

\bibliographystyle{JHEP}
\bibliography{sample}

\end{document}